%% file: ANA-TOPQ-2018-01-PAPER.tex
\pdfoutput=1
\newcommand*{\ATLASLATEXPATH}{}
\documentclass[cernpreprint, texlive=2016, UKenglish]{\ATLASLATEXPATH atlasdoc}
 
\usepackage[subcaption]{\ATLASLATEXPATH atlaspackage}
\usepackage{physics}
\usepackage[capitalize]{cleveref}
\usepackage{\ATLASLATEXPATH atlasbiblatex}
 
\usepackage[unit=false, hepparticle=true, hepprocess=true, jetetmiss=true, misc=true]{\ATLASLATEXPATH atlasphysics}
 
\graphicspath{{logos/}{figs/}}
 
\newif\ifPrintNotes
\PrintNotesfalse  
 
\usepackage{ANA-TOPQ-2018-01-PAPER-defs}

\AtlasTitle{Observation of the associated production of a top quark and a $Z$ boson in $pp$ collisions at $\sqrt{s} = 13\,\text{TeV}$ with the ATLAS detector}

\AtlasAbstract{
Single top-quark production in association with a $Z$ boson, where the $Z$ boson decays to a pair of charged leptons, is measured in the trilepton channel.
The proton--proton collision data collected by the ATLAS experiment from 2015 to 2018
at a centre-of-mass energy of $13\,\text{TeV}$ are used,
corresponding to an integrated luminosity of $139\,\text{fb}^{-1}$.
Events containing three isolated charged leptons (electrons or muons) and
two or three jets, one of which is identified as containing a $b$-hadron, are selected.
The main backgrounds are from $t\bar{t}Z$ and diboson production.
Neural networks are used to improve the background rejection and extract the signal.
The measured cross-section for $t \ell^{+}\ell^{-} q$ production,
including non-resonant dilepton pairs with $m_{\ell^+\ell^-} > 30\,\text{GeV}$,
is $97 \,\pm 13\,\text{(stat.)}\,\pm 7\,\text{(syst.)}\,\text{fb}$,
consistent with the Standard Model prediction.
}
 
\author{The ATLAS Collaboration}
 
\AtlasRefCode{TOPQ-2018-01}
 
\PreprintIdNumber{CERN-EP-2019-273}

\AtlasJournal{JHEP}
\AtlasJournalRef{JHEP 07 (2020) 124}
\AtlasDOI{10.1007/JHEP07(2020)124}

\hypersetup{pdftitle={ATLAS document},pdfauthor={The ATLAS Collaboration}}
 
\begin{document}
 
\maketitle

\tableofcontents
 
 
\section{Introduction}
\label{sec:intro}
This paper reports on the observation of the electroweak production of a top quark or an anti-top quark
associated with a \PZ boson (\tZq) by the ATLAS Collaboration
using a data sample corresponding to an integrated luminosity of \lumi of
proton--proton (\pp) collisions at a centre-of-mass energy of $\sqrt{s} = \SI{13}{\TeV}$.
The final state is where the
\PZ boson decays into electrons or muons, and the \PW boson, from the top-quark decay,
decays into an electron or muon and an associated neutrino, and also includes the contribution from $\tau$-lepton decays into electrons or muons.
The requirement of three leptons in the final state maximises the signal significance relative to the backgrounds.
Evidence for this process has previously been reported by the ATLAS Collaboration~\cite{TOPQ-2016-14}
with a significance of 4.2 standard deviations using data collected in 2015 and 2016.
The CMS Collaboration~\cite{CMS-TOP-18-008} reported an observation with a measured cross-section uncertainty of 15\%
using data collected in 2016 and 2017.
 
\begin{figure}[t!]
\centering
\begin{subfigure}[b]{.30\textwidth}
\includegraphics[width=\textwidth]{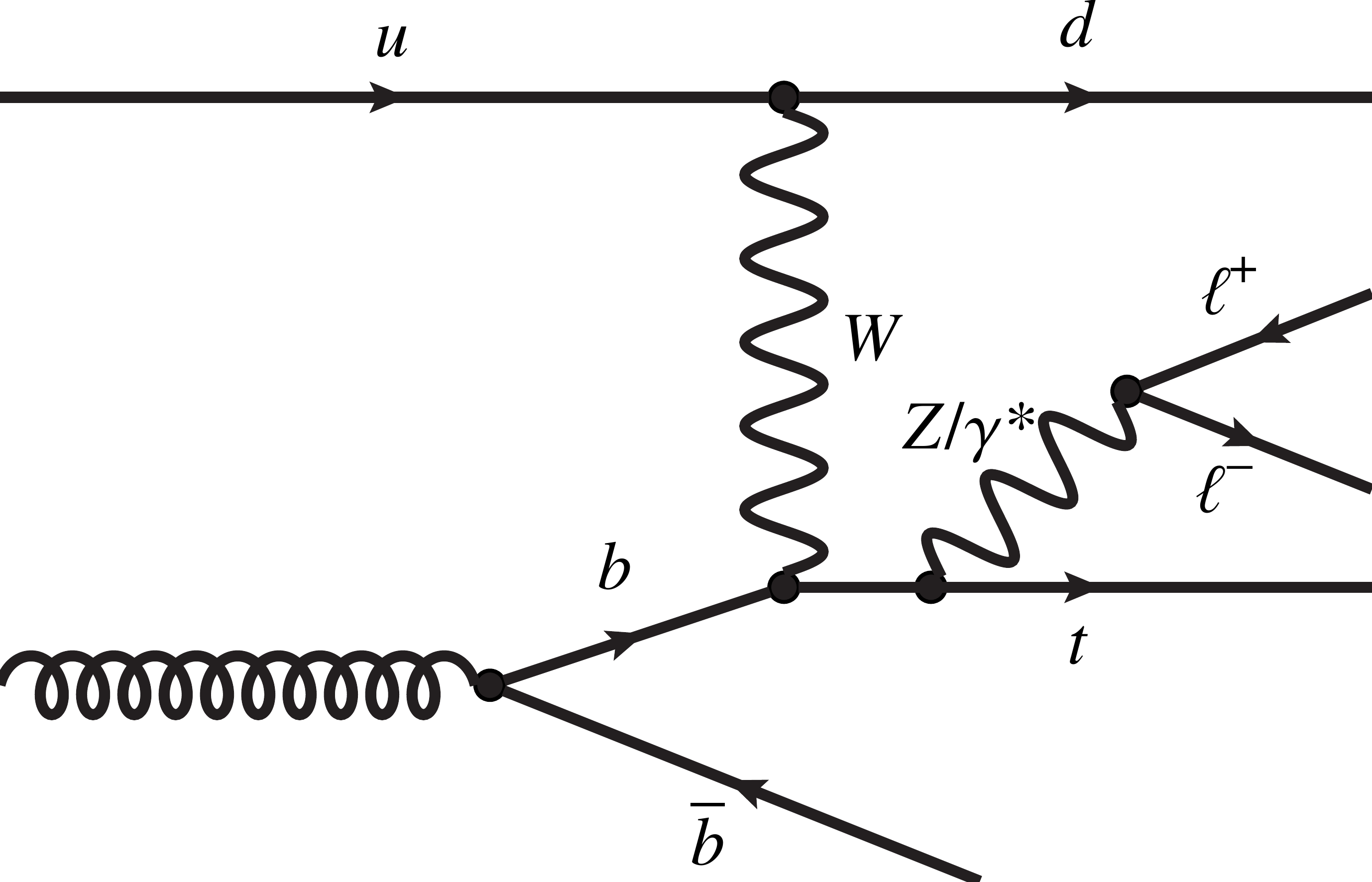}
\caption{}
\label{fig:feynman:a}
\end{subfigure}
\quad
\quad
\begin{subfigure}[b]{.30\textwidth}
\includegraphics[width=\textwidth]{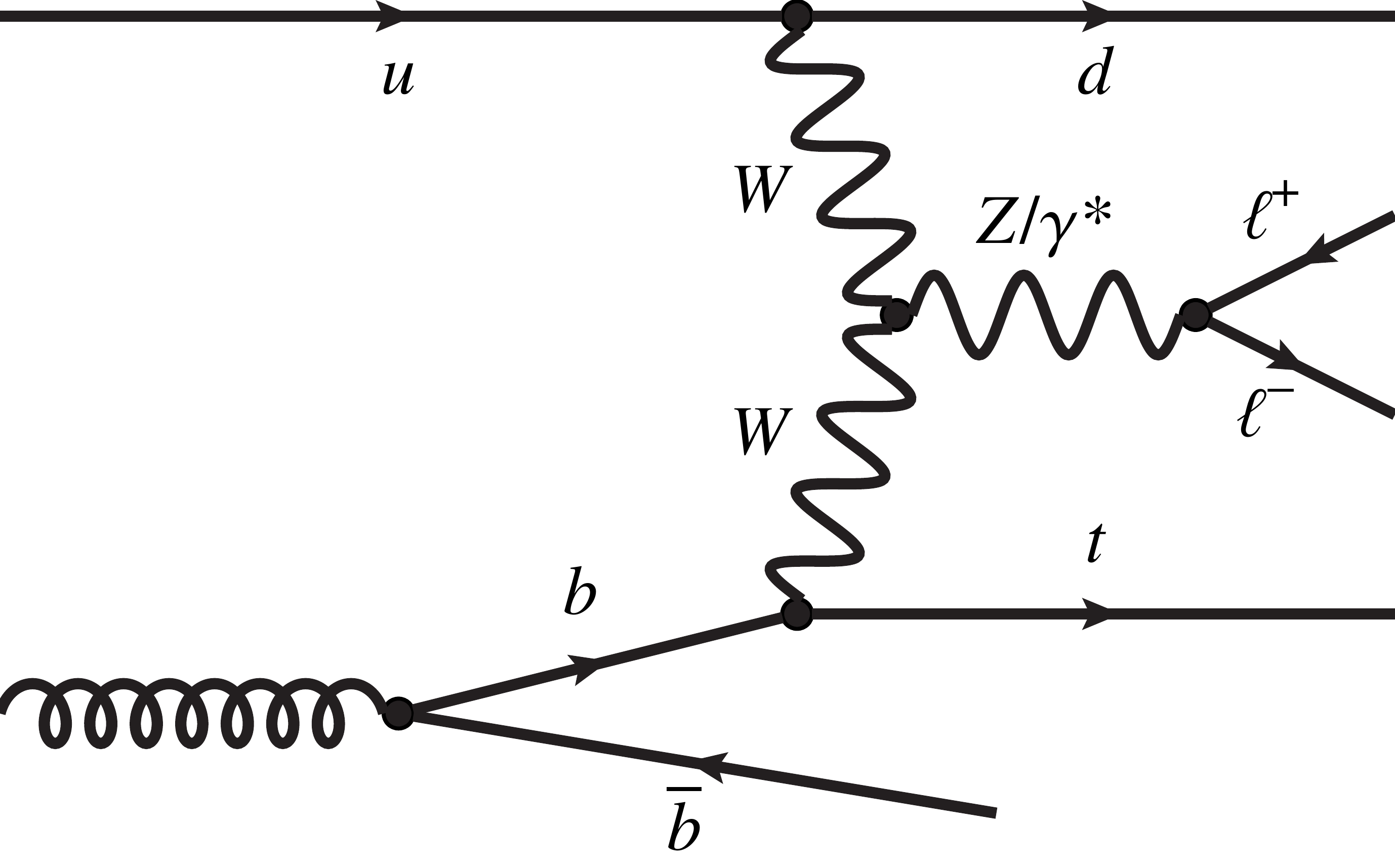}
\caption{}
\label{fig:feynman:b}
\end{subfigure}
\quad
\quad
\begin{subfigure}[b]{.30\textwidth}
\includegraphics[width=\textwidth]{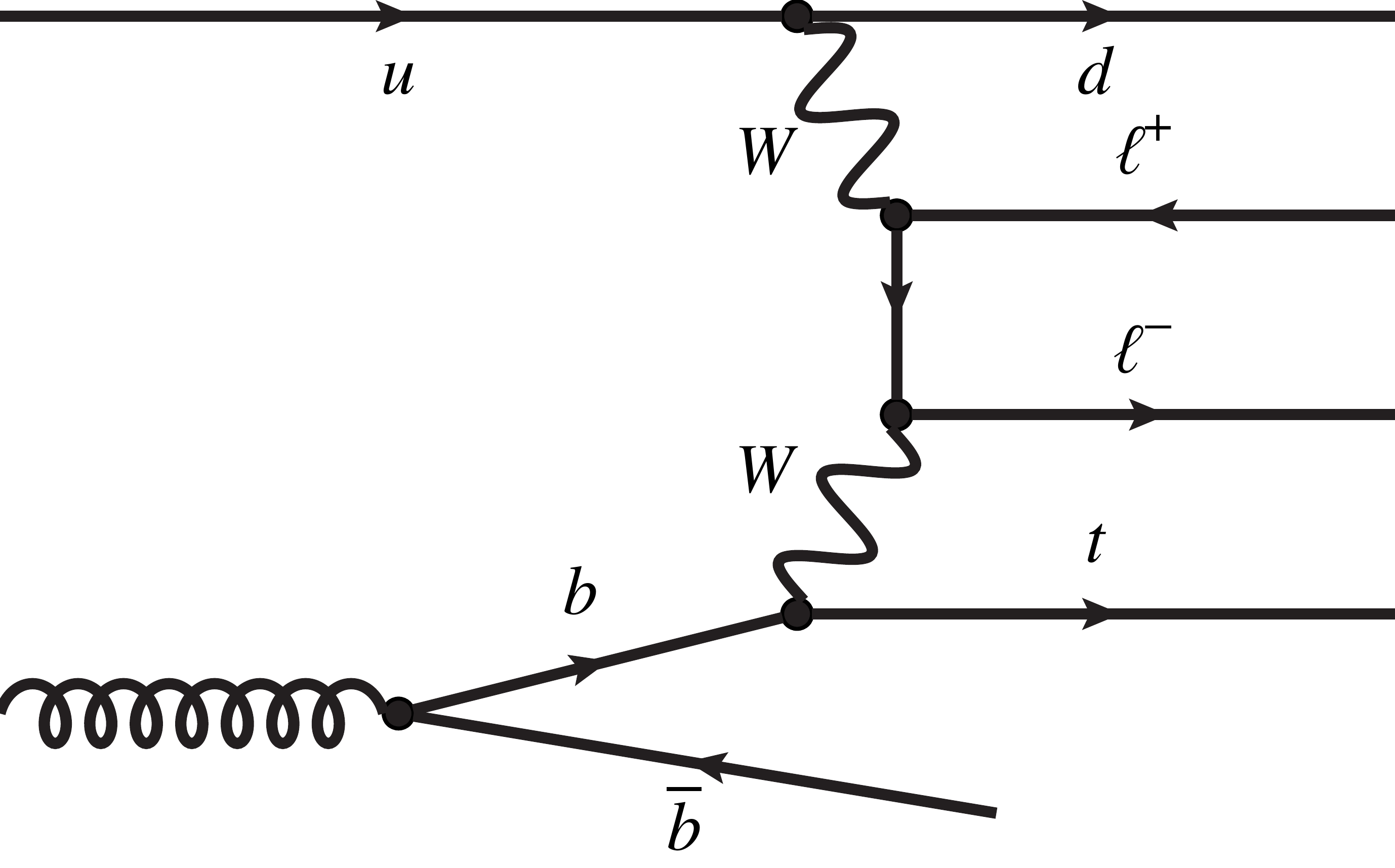}
\caption{}
\label{fig:feynman:c}
\end{subfigure}
\caption{Example Feynman diagrams of the lowest-order amplitudes for the \tZq process, corresponding to (a, b) resonant \Plp{}\Plm production and (c) non-resonant \Plp{}\Plm production. In the four-flavour scheme, the \Pqb-quark originates from gluon splitting.}
\label{fig:feynman}
\end{figure}

At leading order (LO) in the Standard Model (SM) both the single top-quark production and decay occur through the electroweak interaction.
The main LO Feynman diagram is the same as in $t$-channel single top-quark production
with the addition of a \PZ boson radiated from any of the quarks (\Cref{fig:feynman:a})  or
from the $t$-channel \PW-boson propagator (\Cref{fig:feynman:b}).
This allows the \Pqt{}-\PZ
and the \PW{}-\PZ couplings to be indirectly studied in a single interaction.
At LO the \ttZ process is $O(\alpha_\textrm{s}^2)$ in QCD, and the extraction of the \Pqt{}-\PZ coupling is more sensitive to higher-order QCD corrections.
Furthermore, for the \tZq process the next-to-leading-order (NLO) QCD corrections are
small and therefore deviations from the SM can easily be studied
in the framework of the SM effective field theory~\cite{Degrande:2018fog}.
 
In addition to resonant \PZ-boson production,
a small non-resonant \Plp{}\Plm (with \Pl = \Pe, \Pgm, \Pgt) contribution to this process (\tllq)
is accounted for (\Cref{fig:feynman:c}).
Throughout this paper, single top-quark production
with either resonant or non-resonant \Plp{}\Plm in the final state is referred to
as \tZq.
In the SM, the expected cross-section for this process,
calculated at NLO in QCD for a dilepton mass greater than \SI{30}{\GeV}, is $\SI[parse-numbers=false]{102^{+5}_{-2}}{\fb}$.

 
\section{ATLAS detector}
\label{sec:atlas}
\newcommand{\AtlasCoordFootnote}{
ATLAS uses a right-handed coordinate system with its origin at the nominal interaction point (IP)
in the centre of the detector and the $z$-axis along the beam pipe.
The $x$-axis points from the IP to the centre of the LHC ring,
and the $y$-axis points upwards.
Cylindrical coordinates $(r,\phi)$ are used in the transverse plane,
$\phi$ being the azimuthal angle around the $z$-axis.
The pseudorapidity is defined in terms of the polar angle $\theta$ as $\eta = -\ln \tan(\theta/2)$.
Distances in the $\eta$--$\phi$ plane are measured in units of $\Delta R \equiv \sqrt{(\Delta\eta)^{2} + (\Delta\phi)^{2}}$.}

The ATLAS detector~\cite{PERF-2007-01} at the LHC covers nearly the entire solid angle around the collision point.\footnote{\AtlasCoordFootnote}
It consists of an inner tracking detector surrounded by a thin superconducting solenoid, electromagnetic and hadronic calorimeters,
and a muon spectrometer incorporating three large superconducting toroidal magnets.
 
The inner-detector system (ID) is immersed in a \SI{2}{\tesla} axial magnetic field
and provides charged-particle tracking in the range \(|\eta| < 2.5\).
The high-granularity silicon pixel detector covers the vertex region and typically provides four measurements per track,
the first hit being normally in the insertable B-layer installed before Run~2~\cite{ATLAS-TDR-2010-19,PIX-2018-001}.
It is followed by the silicon microstrip tracker, which usually provides eight measurements per track.
These silicon detectors are complemented by the transition radiation tracker (TRT),
which enables radially extended track reconstruction up to \(|\eta| = 2.0\).
The TRT also provides electron identification information
based on the fraction of hits (typically 30 in total) above a higher energy-deposit threshold corresponding to transition radiation.
 
The calorimeter system covers the pseudorapidity range \(|\eta| < 4.9\).
Within the region \(|\eta|< 3.2\), electromagnetic (EM) calorimetry is provided by barrel and
endcap high-granularity lead/liquid-argon (LAr) calorimeters,
with an additional thin LAr presampler covering \(|\eta| < 1.8\),
to correct for energy loss in material upstream of the calorimeters.
Hadronic calorimetry is provided by the steel/scintillator-tile calorimeter,
segmented into three barrel structures within the region \(|\eta| < 1.7\), and two copper/LAr hadronic endcap calorimeters.
The solid angle coverage is completed with forward copper/LAr and tungsten/LAr calorimeter modules
optimised for electromagnetic and hadronic measurements respectively.
 
The muon spectrometer comprises separate trigger and
high-precision tracking chambers measuring the deflection of muons in a magnetic field generated by superconducting air-core toroids.
The field integral of the toroids ranges between \num{2.0} and \SI{6.0}{\tesla\metre}
across most of the detector.
A set of precision chambers covers the region \(|\eta| < 2.7\) with three layers of monitored drift tubes,
complemented by cathode-strip chambers in the forward region, where the background is highest.
The muon trigger system covers the range \(|\eta| < 2.4\) with resistive-plate chambers in the barrel and thin-gap chambers in the endcap regions.
 
Interesting events are selected for recording by the first-level trigger system implemented in custom hardware,
followed by selections made by algorithms implemented in software in the high-level trigger~\cite{TRIG-2016-01}.
The first-level trigger selects events from the \SI{40}{\MHz} bunch crossings at a rate below \SI{100}{\kHz},
which the high-level trigger further reduces to record events to disk at about \SI{1}{\kHz}.
 
 
 
\section{Data and simulation samples}
\label{sec:datamc}
 
The data sample used in this article corresponds to \lumi of \pp collisions at \(\sqrt{s} = \SI{13}{\TeV}\)
collected by the ATLAS detector during 2015--2018, after requiring stable LHC beams and that all detector subsystems were operational~\cite{DAPR-2018-01}.
 
Candidate events were required to satisfy one of the single-electron triggers or one of the single-muon
triggers~\cite{TRIG-2016-01,TRIG-2018-05-INT,ATL-DAQ-PUB-2016-001,ATL-DAQ-PUB-2017-001,ATL-DAQ-PUB-2018-002,ATL-DAQ-PUB-2019-001}.
Single-lepton triggers with low transverse momentum, \pT, thresholds and standard isolation requirements were
combined in a logical OR with higher-threshold triggers that had a looser identification criterion and
did not have any isolation requirement, resulting in an efficiency of almost 100\% for events passing the analysis selection.
The lowest \pT threshold used for electrons was \SI[parse-numbers=false]{24}{\GeV} (\SI[parse-numbers=false]{26}{\GeV}) in 2015 (2016--2018),
while for muons the threshold was \SI[parse-numbers=false]{20}{\GeV} (\SI[parse-numbers=false]{26}{\GeV}).
 
To evaluate the effects of the detector resolution and acceptance on
the signal and background, and to estimate the SM backgrounds, simulated event samples were produced using a $\GEANT4$-based Monte Carlo (MC) detector simulation~\cite{Agostinelli:2002hh-INT, SOFT-2010-01}. Some of the samples used for evaluating systematic uncertainties did not use the full $\GEANT4$ simulation but instead relied on parameterised showers in the calorimeter.
The top-quark mass in the event generators described below was set to \SI{172.5}{\GeV}.
 
The simulated data must account for the fact that significantly more than one inelastic \pp collision occurs per bunch
crossing.
The average number of collisions per bunch crossing ranged from 13 to 38
for the 2015 through 2018 data-taking periods,
respectively. Inelastic
collisions were simulated using \PYTHIAV{8.186}~\cite{Sjostrand:2007gs} with the A3 set of tuned parameters~\cite{ATL-PHYS-PUB-2016-017} and the NNPDF2.3~LO~\cite{Ball:2012cx} set of parton distribution functions (PDFs), and overlaid on the
signal and background MC samples.
These simulated events were reweighted to match the conditions of the collision data, specifically the additional \pp interactions (\pileup).
 
To estimate the signal acceptance and efficiency, and to study the effect of
different selection criteria on the expected precision of the measurement,
a \tZq sample was simulated, including non-resonant \Plp{}\Plm contributions.
The sample was generated using the four-flavour scheme at NLO in QCD with \MGMCatNLOV{2.6.0}~\cite{Alwall:2014hca},
requiring the dilepton invariant mass to be larger than \SI{30}{\GeV} and
using the \textsc{NNPDF30\_nlo\_as\_0118\_nf\_4}~\cite{Ball:2014uwa-INT}
PDF set. Following the discussion in Ref.~\cite{Frederix:2012dh},
the functional form of the renormalisation and factorisation scale was set to $4\sqrt{m^{2}_{\Pqb} + p^{2}_{\text{T},\Pqb}}$,
where the \Pqb-quark is the one produced in the gluon splitting in the initial state associated with \tZq production.
The parton showering and hadronisation in signal events were simulated
using \PYTHIAV{8.230}~\cite{Sjostrand:2014zea}, with a set of tuned parameters selected according to
Ref.~\cite{ATL-PHYS-PUB-2014-021}, referred to as the \enquote{A14 tune}.
 
The predicted cross-section was calculated with \MGMCatNLOV{2.6.0}, using the five-flavour scheme with the \textsc{NNPDF30\_nlo\_as\_0118} PDF set and with the renormalisation and factorisation scales, \muR\ and \muF, set to $\muR = \muF = (\mtop + \mZ)/4 = \SI{66}{\GeV}$. The SM \tZq cross-section at NLO in QCD, including non-resonant contributions with $m_{\ell^+\ell^-} > 30\,\text{GeV}$, is \SI{102}{\fb}.
The renormalisation and factorisation scale uncertainties are $^{+5.2}_{-1.3}\%$ and the PDF uncertainty is $\pm1.0\%$. \LDnote{}{Muhammad to add how scale unc is obtained}
The PDF uncertainty was calculated
using the replica method described in Ref.~\cite{Butterworth:2015oua}.
 
The background to the signal is estimated by using simulated samples
that contain at least two leptons and at least two jets.
These samples include the production of \ttbar,
\ttH, \ttZ, \ttW, \tW, \tWZ,
diboson (\PW{}\PW, \PW{}\PZ, or \PZ{}\PZ), and \Zjets events.
 
The nominal \ttbar and \ttH simulated samples were generated using the
NLO matrix-element generator
\POWHEGBOXV{v2}~\cite{Nason:2004rx, Frixione:2007vw, Alioli:2010xd,Frixione:2007nw,Hartanto:2015uka} with the
parameter \hdamp, which controls the transverse momentum of the first additional gluon emission beyond the Born configuration, 
set to $1.5 \times \mtop$ for \ttbar~\cite{ATL-PHYS-PUB-2016-020} and $0.75 \times (2\times\mtop+ m_{\PH})$ for \ttH,
with $\mH = \SI{125}{\GeV}$.
These events were then passed through \PYTHIAV{8.230}
to generate the underlying event and perform the parton showering and
hadronisation.
The PDF set used in the sample simulation was NNPDF3.0~NLO
and for \PYTHIAV{8} the A14 tune and NNPDF2.3~LO PDF set were used.
 
Additional \ttbar simulated samples are used to assess modelling uncertainties~\cite{ATL-PHYS-PUB-2018-009}.
To evaluate the uncertainty due to initial-state radiation, samples with higher parton radiation were produced
by decreasing the factorisation and renormalisation scales by a factor of 0.5 and simultaneously increasing the \hdamp parameter to twice its nominal value, and using the \enquote{Var3c} up variation from the A14 tune~\cite{ATL-PHYS-PUB-2017-007}. For lower parton radiation, the nominal \hdamp value was used, while the renormalisation and factorisation scales were increased by a factor of two and the \enquote{Var3c} down variation was selected in the parton shower.
To study the impact of using an alternative parton shower and hadronisation model, a sample was produced with
the \POWHEGBOXV{v2} generator interfaced to \HERWIGV{7.0.4}~\cite{Bahr:2008pv-INT,Bellm:2015jjp}, the former using the NNPDF3.0~NLO PDF set and the latter using the H7UE set of tuned parameters~\cite{Bellm:2015jjp} and the MMHT2014~LO PDF set~\cite{Harland-Lang:2014zoa}.
To assess the uncertainty due to the choice of matching scheme, a
sample generated by \MGMCatNLOV{2.6.0} with the NNPDF3.0~NLO PDF set was passed through \PYTHIAV{8.230}, which used the A14 tune and NNPDF2.3\ LO PDF set.
For all samples produced with this set of generators, the matrix-element correction for the first emission was turned off and the global recoil option was used.
 
The production of \ttZ and \ttW events was modelled using the \MGMCatNLOV{2.3.3} generator at NLO in QCD, with the NNPDF3.0~NLO PDF set.
Parton showering and hadronisation were modelled with \PYTHIAV{8.210}, using the A14 tune and the NNPDF2.3~LO PDF set.
Non-resonant \Plp{}\Plm contributions are included for \ttZ.
 
To assess modelling uncertainties, alternative \ttZ simulated samples were produced with the \SHERPAV{2.2.1}~\cite{Gleisberg:2008ta-INT} generator
with one additional parton at NLO accuracy.
The CKKW matching scale of the
additional emissions was set to \SI{30}{\GeV}.
The default \SHERPAV{2.2.1} parton shower was used along with the NNPDF3.0~NNLO PDF set.
 
The \tW simulated samples used the same generators as those used for the various \ttbar samples~\cite{Re:2010bp}.
To avoid overlap between \tW and \ttbar production, the diagram removal (DR) scheme was employed~\cite{Frixione:2008yi-INT},
where all NLO diagrams that overlap with the doubly resonant \ttbar
contributions are removed from the calculation of the \tW amplitude.
 
The production of \tWZ events was modelled using the \MGMCatNLOV{2.3.3}
generator at NLO with the NNPDF3.0~NLO PDF set.
The generator was interfaced to \PYTHIAV{8.212}, which used the A14 tune and the NNPDF2.3~LO PDF set.
The modelling uncertainties for this process are evaluated by comparing with a MC sample produced using the same generators but employing a different treatment of the interference between \ttZ and \tWZ, namely the diagram removal scheme that takes the interference term into account (DR2), as opposed to the nominal DR1 scheme~\cite{Demartin:2016axk}.
 
The simulation of the diboson event samples used the NLO \SHERPA event generators:
\SHERPAV{2.2.1} for events with one boson decaying into hadrons,
and \SHERPAV{2.2.2} for events with both bosons decaying into leptons.
In this set-up, multiple
matrix elements were matched and merged with the \SHERPA parton shower
based on Catani--Seymour dipole factorisation~\cite{Gleisberg:2008fv,Schumann:2007mg}
using the MEPS@NLO
prescription~\cite{Hoeche:2011fd,Hoeche:2012yf-INT,Catani:2001cc,Hoeche:2009rj}.
The simulations included up to one additional parton at NLO accuracy
and up to three additional parton emissions at LO accuracy.
The virtual QCD corrections for matrix elements at NLO accuracy were
provided by the \openloops library~\cite{Cascioli:2011va-INT,Denner:2016kdg}
with the NNPDF3.0~NLO PDF set.
 
To assess modelling uncertainties for the diboson background, the
\POWHEGBOXV{v2}~\cite{Nason:2013ydw} generator was used to generate the diboson processes at NLO in QCD\@.
Events were generated using the CT10\ NLO PDF set~\cite{Lai:2010vv} and showered with \PYTHIAV{8.210}
with the AZNLO~\cite{STDM-2012-23} tune and the CTEQ6L1~\cite{Pumplin:2002vw} PDF set.
 
For the modelling of \Zjets, a sample was
generated using \SHERPAV{2.2.1} with the NNPDF3.0~NNLO PDF set. The matching and merging procedure, as well as the virtual QCD corrections, were similar to those described for the diboson event simulation.
The simulations included up to two additional partons at NLO accuracy and up to four additional parton emissions at LO accuracy.
 
 
\section{Object reconstruction}
\label{sec:reco}
 
The reconstruction of the basic objects used in the analysis is described in the following.
The primary vertex~\cite{ATL-PHYS-PUB-2015-026} is selected as the \pp vertex candidate with the highest sum of
the squared transverse momenta of all associated tracks with $\pT > \SI{400}{\MeV}$.
 
Electron candidates are reconstructed from energy clusters in the
EM calorimeter that match a reconstructed track~\cite{EGAM-2018-01-INT}.
The clusters are required to be within the range $\abseta < 2.47$,
excluding the transition region between the barrel and endcap calorimeters at
$1.37 < \abseta < 1.52$.
Electron candidates must also satisfy a
transverse energy requirement of $\ET > \SI{20}{\GeV}$~\cite{EGAM-2018-01-INT}.
A likelihood-based discriminant is constructed from a set of
variables that enhance the electron selection, while
rejecting photon conversions and hadrons misidentified as
electrons~\cite{EGAM-2018-01-INT}.
An $\eta$- and \ET-dependent
selection on the likelihood discriminant is applied, such that it
has an 80\% efficiency when used to identify electrons from \PZ{}-boson decays.
Electrons are further required to be isolated using criteria based on ID tracks and topological clusters in the calorimeter,
with an efficiency of 90\% (99\%) for $\ET = \SI[parse-numbers=false]{25}{\GeV}~(\SI[parse-numbers=false]{60}{\GeV})$.
Correction factors are applied to simulated electrons to take into account the small differences in
reconstruction, identification and isolation efficiencies between data and MC simulation.
 
Muon candidates are reconstructed
by combining a reconstructed track from the inner detector with one
from the muon spectrometer~\cite{PERF-2015-10}, and are required to have $\pt>\SI{20}{\GeV}$
and $\abseta<2.5$.
To reject misidentified
muon candidates, primarily originating from pion and kaon decays, several
quality requirements are imposed on the muon candidate.
An isolation requirement based on ID tracks and topological clusters in the calorimeter is imposed,
resulting in an efficiency of 90\% (99\%) for $\pT = \SI[parse-numbers=false]{25}{\GeV}~(\SI[parse-numbers=false]{60}{\GeV})$.
The overall efficiency obtained for muons from \PW-boson decays in simulated \ttbar events is 96\%.
Like for electrons, correction factors are applied to simulated muons to account for the small differences between data and simulation.
 
Jet reconstruction in the calorimeter starts from topological clustering~\cite{PERF-2014-07} of individual calorimeter cells calibrated to the electromagnetic energy scale. The \antikt algorithm~\cite{Cacciari:2005hq,Cacciari:2008gp}, with the radius parameter set to $R=0.4$, is used to reconstruct the jets~\cite{Fastjet}. These jets are then calibrated to the particle level by the application of a jet energy scale derived from simulation and in situ corrections based on $\sqrt{s}=\SI{13}{\TeV}$ data~\cite{PERF-2016-04}.
Jets are required to have $\pT > \SI{35}{\GeV}$ in the region $\abseta < 4.5$.
To suppress jets arising from \pileup, a discriminant called the \enquote{jet vertex tagger} (JVT) is constructed using a two-dimensional likelihood method~\cite{PERF-2014-03}.
For jets with $\pT < \SI{60}{\GeV}$ and $\abseta < 2.5$
a JVT requirement corresponding to a 92\% efficiency, while rejecting 98\% of jets from \pileup and noise, is imposed.
To reject jets at high $\abseta$ originating from additional \pp interactions,
a forward jet vertex tagger (fJVT) requirement is applied~\cite{PERF-2016-06}.
All jets with $\abseta > 2.5$ are required to satisfy the requirements of the fJVT \enquote{medium} working point.
This has an efficiency of selecting hard-scattered jets of up to 97\%
and a \pileup-jet efficiency of 53\% for jets with a \pT of \SI{40}{\GeV}.
 
To identify jets containing a \Pqb-hadron (\bjets), a multivariate algorithm
is employed~\cite{FTAG-2018-01-INT,ATL-PHYS-PUB-2017-013}.
It uses impact parameter and reconstructed secondary vertex information from
tracks contained in the jet as input.
A calibration in bins of \pT is derived at four efficiency points.
Each jet is assigned a score, depending on how many of the efficiency points are passed.
Due to its use of the ID, the reconstruction of \bjets is restricted to the region $\abseta < 2.5$.
Candidate \Pqb-jets must have a \btag discriminant value that exceeds a threshold such that a 70\% \bjet
selection efficiency is achieved in simulated \ttbar events.
With this criterion,
the misidentification rate for light-jets, i.e.\ jets containing neither a \Pqb- nor a \Pqc-hadron, is 0.3\%, while it is 11\% for jets initiated by \Pqc-quarks.
Correction factors are derived and applied to correct for differences in \bjet selection efficiency and the mistagging rates between data and MC simulation~\cite{FTAG-2018-01-INT,ATLAS-CONF-2018-001,ATLAS-CONF-2018-006}.
 
The missing transverse momentum, with magnitude \met, is calculated as the negative of the vector sum of
the transverse momenta of all reconstructed objects.
To account for the soft hadronic activity, a term including tracks associated with the primary vertex but not with any of the reconstructed objects is added to the \MET calculation~\cite{PERF-2016-07}.
 
To avoid cases where the detector response to a single physical object is reconstructed as two separate final-state objects, an overlap removal procedure is used.
If electron and muon candidates share a track, the electron candidate is removed. After that, if the $\DeltaR_{y,\phi}$ distance\footnote{$\DeltaR_{y,\phi}$ is the Lorentz-invariant distance in the rapidity--azimuthal-angle plane, defined as $\DeltaR_{y,\phi} =\sqrt{(\Delta y)^2+(\Delta\phi)^2}$.} between a jet and an electron candidate is less than 0.2, the jet is discarded. If multiple jets satisfy this requirement, only the closest jet is removed.
For jet--electron distances between 0.2 and 0.4, the electron candidate is removed.
If the distance
between a jet and a muon candidate is less than 0.4, the muon candidate is removed if the jet has more
than two associated tracks, otherwise the jet is removed.
 
 
\section{Signal and control regions}
\label{sec:evtsel}
 
The \tZq final state used for this measurement comprises three
charged leptons (electrons or muons), missing transverse momentum,
one \bjet from the top-quark decay and
an additional jet that does not satisfy the \btag requirement (untagged jet) and is expected to be emitted preferentially at high $\abseta$.
A second untagged jet is allowed in order to include events with QCD radiation.
To help separate the \tZq signal from the backgrounds that do not contain a $Z$ boson and a top quark
(diboson, \Zjets and \ttbar events), both the \PZ-boson and
the top-quark invariant masses are reconstructed.
 
The data are divided into eight non-overlapping regions: two signal regions (SR)
designed to select \tZq events and six control regions (CR)
designed to enhance the selection of the
main sources of background events (\ttZ, diboson, \Zjets and \ttbar events).
The CRs are used to adjust the normalisation and reduce the associated systematic uncertainties in the main backgrounds.
 
\Cref{tab:cutsummary} summarises the selection criteria applied across all the regions considered.
Leptons and jets have to satisfy the requirements discussed in~\cref{sec:reco} and one of the
leptons is required to have $\pT > \SI{28}{\GeV}$, because of the trigger thresholds, and match, with $\Delta R < 0.15$, the lepton reconstructed by the trigger.
The nomenclature \textit{n}j\textit{m}b is used to denote the regions, where \textit{n} is the total number of jets, of which \textit{m} are \Pqb-tagged.
 
The SRs require an opposite-sign,
same-flavour (\OSSF) lepton pair to reconstruct the \PZ boson.
In the \mee and \emm channels the pair is uniquely identified, whereas
in the \eee and \mmm channels both of the possible combinations are considered and the
pair with the invariant mass closer to the \PZ-boson mass is chosen.
The dilepton invariant mass requirement is $|\mll-\mZ|<\SI{10}{\GeV}$.
The remaining lepton and the \MET are used to reconstruct the leptonically decaying \PW boson.\footnote{
The longitudinal component of the neutrino four-momentum is obtained by using the mass constraint of the \PW boson.
The twofold ambiguity is resolved by choosing the solution with the smaller $|p_z^{\nu}|$, since the \PW boson is expected to be produced with small pseudorapidity~\cite{TOPQ-2011-18}.}
The four-momenta of the reconstructed \PW boson and the \bjet  are summed to reconstruct the top quark.
 
\begin{table}[t!]
\caption{
Overview of the requirements applied when selecting events in the signal and control regions.
OSSF is an opposite-sign same-flavour lepton pair. OSDF is an opposite-sign different-flavour lepton pair.
}
\label{tab:cutsummary}
\begin{center}
\begin{tabular}{cccc}
\toprule
& \multicolumn{2}{c}{Common selections} & \\
\midrule
& \multicolumn{2}{c}{Exactly 3 leptons ($e$ or $\mu$) with $\abseta < 2.5$} &  \\
& \multicolumn{2}{c}{$\pT(\ell_1)> \SI{28}{\GeV}$, $\pT(\ell_2)> \SI{20}{\GeV}$, $\pT(\ell_3)> \SI{20}{\GeV}$} &  \\
& \multicolumn{2}{c}{$\pT(\text{jet})> \SI{35}{\GeV}$} &  \\[0.3em]
\toprule
SR 2j1b   & CR diboson 2j0b & CR \ttbar 2j1b & CR \ttZ 3j2b \\
\midrule
$\ge$ 1 \OSSF pair & $\ge$ 1 \OSSF pair & $\ge$ 1 \OSDF pair & $\ge$ 1 \OSSF pair \\
$|\mll - \mZ| <  \SI{10}{\GeV}$ & $|\mll - \mZ| < $ \SI{10}{\GeV} & No \OSSF pair & $|\mll - \mZ| < $ \SI{10}{\GeV} \\
2 jets, $\abseta < $ 4.5  & 2 jets, $\abseta < $ 4.5  & 2 jets, $\abseta < $ 4.5 & 3 jets, $\abseta < $ 4.5 \\
1 \bjet, $\abseta < $ 2.5 & 0 \bjets  & 1 \bjet, $\abseta < $ 2.5 & 2 \bjets, $\abseta < $ 2.5   \\[0.3em]
\toprule
SR 3j1b   & CR diboson 3j0b & CR \ttbar 3j1b & CR \ttZ 4j2b \\
\midrule
$\ge$ 1 \OSSF pair & $\ge$ 1 \OSSF pair & $\ge$ 1 \OSDF pair & $\ge$ 1 \OSSF pair \\
$|\mll - \mZ| <  \SI{10}{\GeV}$ & $|\mll - \mZ| < $ \SI{10}{\GeV} & No \OSSF pair & $|\mll - \mZ| < $ \SI{10}{\GeV} \\
3 jets, $\abseta < $ 4.5  & 3 jets, $\abseta < $ 4.5  & 3 jets, $\abseta < $ 4.5 & 4 jets, $\abseta < $ 4.5 \\
1 \bjet, $\abseta < $ 2.5 & 0 \bjets  & 1 \bjet, $\abseta < $ 2.5 & 2 \bjets, $\abseta < $ 2.5   \\
\bottomrule
\end{tabular}
\end{center}
\end{table}

The diboson CRs use the same event selection as each of the SRs but a \Pqb-tag
veto is applied.
The CRs for \ttZ are the same as the SRs except for the addition of another \bjet.
The CRs for \ttbar are the same as the
SRs except for the requirement of no OSSF lepton pair
and at least one opposite-sign, different-flavour (OSDF) lepton pair.
 
These SRs and CRs are included in a binned maximum-likelihood fit that is performed to measure the signal cross-section. In the fit, the normalisations of the signal and the \ttbar (including \tW) and \Zjets backgrounds are unconstrained, while the other backgrounds are constrained to be close to their SM prediction (see~\cref{sec:sys}).
 
 
 
\section{Background estimation}
\label{sec:backgr}
 
Two classes of backgrounds are considered: processes in which three or more prompt leptons are produced, such as diboson production or the associated production of a top-quark pair and a boson (\PW, \PZ or \PH); and processes with only two prompt leptons in the final state (such as \Zjets, \ttbar and \tW production) and one additional non-prompt or fake lepton that satisfies the selection criteria.
Such non-prompt or fake
leptons can originate from decays of bottom or charm hadrons, from jets misidentified as electrons, leptons from kaon or pion decays, or electrons
from
photon conversions.
 
After applying the SR event selection,
diboson, \ttZ, \Zjets and \ttbar production constitute the largest
backgrounds. For the SR 2j1b, the dominant background
source is diboson production, mainly \WZ events. Monte Carlo simulation indicates that these represent 50\% of the
total number of selected background events in this region, while the second largest backgrounds, \ttZ and \tWZ amount to 32\%.
Non-prompt-lepton backgrounds are predicted to contribute up to 14\% of the events.
In the SR 3j1b, \ttZ is the dominant
background source, giving 45\% of background events. Diboson production yields
28\% of all background events in that region.
 
The diboson contribution is split according to the origin of the associated jets using generator-level information. If one of the jets contains a \Pqb- or \Pqc-hadron then it is classified as diboson + heavy flavour (\VVHF), otherwise the event is classified as diboson + light flavour (\VVLF).
 
The \ttZ and \tWZ backgrounds are combined, as are \ttbar and \tW.
The \ttW and \ttH contributions are also combined, since both are very small.
 
All background contributions involving prompt leptons are estimated by using MC
samples that are normalised to their respective SM predicted cross-sections
calculated at NLO in QCD.
The cross-section of the \ttH background includes NLO+NLL soft-gluon resummation~\cite{Kulesza:2015vda}.
 
The estimation of the non-prompt-lepton background using MC samples is challenging. The simulation does not accurately model the rate for prompt-lepton misidentification. In addition, it suffers from low statistics after applying the SR event selection requirements, leading to large fluctuations in the predicted event kinematics. The first issue is addressed by normalising the non-prompt-lepton predictions to data in dedicated CRs. The method developed to address the second issue is discussed in the following.
Generator-level studies have shown that for \ttbar events\footnote{
This is also valid for the very small number of \tW events. The same method of estimating their contribution is used for both processes.}
and \Zjets events passing the event selection, the additional non-prompt lepton usually originates from a charm- or bottom-hadron decay, with a smaller contribution from other sources, such as photon conversions in the case of non-prompt electrons.

Therefore, the method used to estimate the shape of kinematic distributions for non-prompt-lepton backgrounds with high statistical precision uses MC samples enriched in events with semileptonic \bjet decays.
Generated events are used for \ttbar + \tW and \Zjets independently,
with a preselection of two leptons instead of three leptons and two \bjets.
One of the two \bjets, which is selected at random,
is replaced by a lepton. 
The energy and polar angle of the replacement lepton, relative to the direction of
the \bjet momentum, are derived from a generator-level study of the polar-angle distribution versus lepton energy
in the rest frame of the \Pqb-hadron, which is assumed to carry the \bjet energy.
The azimuthal angle is generated uniformly around the \bjet direction.
The lepton four-momenta are transformed to the laboratory frame using the \bjet four-momenta.
If the \bjet is within a cone of $\DeltaR = 0.4$ around the lepton, the \bjet is removed.
In the other events the \bjet is kept and the
event will ultimately not satisfy the common selection, as it contains two \Pqb-jets.
The distributions of kinematic variables from the resulting sample are compared with those from the \ttbar + \tW and \Zjets MC samples, applying the common selection, and are found to agree within statistical uncertainties.

 
 
\section{Multivariate analysis}
\label{sec:nn}
To improve the separation between signal and background, a 
neural network (NN) is used to derive a discriminant that takes
correlations among the input variables into account.
The package NeuroBayes~\cite{Feindt:2004wla,Feindt:2006pm} is used,
which combines a three-layer feed-forward NN with a complex, robust
preprocessing that orders the input variables by their separation power~\cite{TOPQ-2011-14}.
The first layer of the network consists of one input node for each input variable plus one bias node~\cite{Feindt:2004wla}.
The second layer can have an arbitrary, user-defined, number of hidden nodes.
There is one output node that gives a continuous value in the interval $[-1,1]$.
The NN uses Bayesian regularisation~\cite{Feindt:2004wla} during the training process
to improve its performance and stability, and to avoid overtraining.
All background processes are considered in the training and are weighted according
to the expected number of events in each SR.
 
Only variables that provide good separation and are well modelled are used
in the final NN\@.
A separate network is trained for each of the SRs, with each NN starting
with the same input variables and 25 hidden nodes.
The 15 input variables that
give the best separation according to NeuroBayes are used for the final NN
training. The full list of the variables used for the NN training is shown in
\cref{tab:inputvars}.
The untagged jet is denoted \jf. When two untagged jets are selected, \jf (\jr) refers to the one for which the invariant mass of this untagged jet and the \Pqb-tagged jet is the largest (smallest).
 
Variables related to the reconstructed \PZ boson
help to reduce the \ttbar background, while top-quark-related quantities are useful in separating the signal from processes such as \WZ and \Zjets,
in which no top quark is produced.
For the signal, the untagged jet comes from the spectator quark in the hard-scattering
process and thus tends to have higher $\abseta$, helping the NN provide better separation
from diboson and \ttZ background events.
The invariant mass of an untagged jet and \bjet, \mbj, is also used
as a discriminating variable for selecting signal over background.
 
In terms of the separation power, the four highest-ranked variables for
the two SRs are the same. The highest-ranked variable is \mbj.
The \mbj value is larger for topologies with a forward jet, because of the large angular separation between the forward jet and the central \bjet.
The other three variables are the reconstructed top-quark mass,
the absolute value of the untagged \jf jet pseudorapidity,
and the transverse mass of the \PW boson.\footnote{
The transverse mass is calculated using the momentum of the lepton associated with
the \PW boson, \MET and the azimuthal angle, $\phi$, between the two:
$m_{\text{T}}\left(\ell, \MET\right) = \sqrt{2 \pT(\ell) \MET
\left[1-\cos \Delta \phi \left(\ell, \MET \right) \right]}$.
}
 
\begin{figure}[htbp]
\centering
\begin{subfigure}[b]{.32\textwidth}
\includegraphics[width=\textwidth]{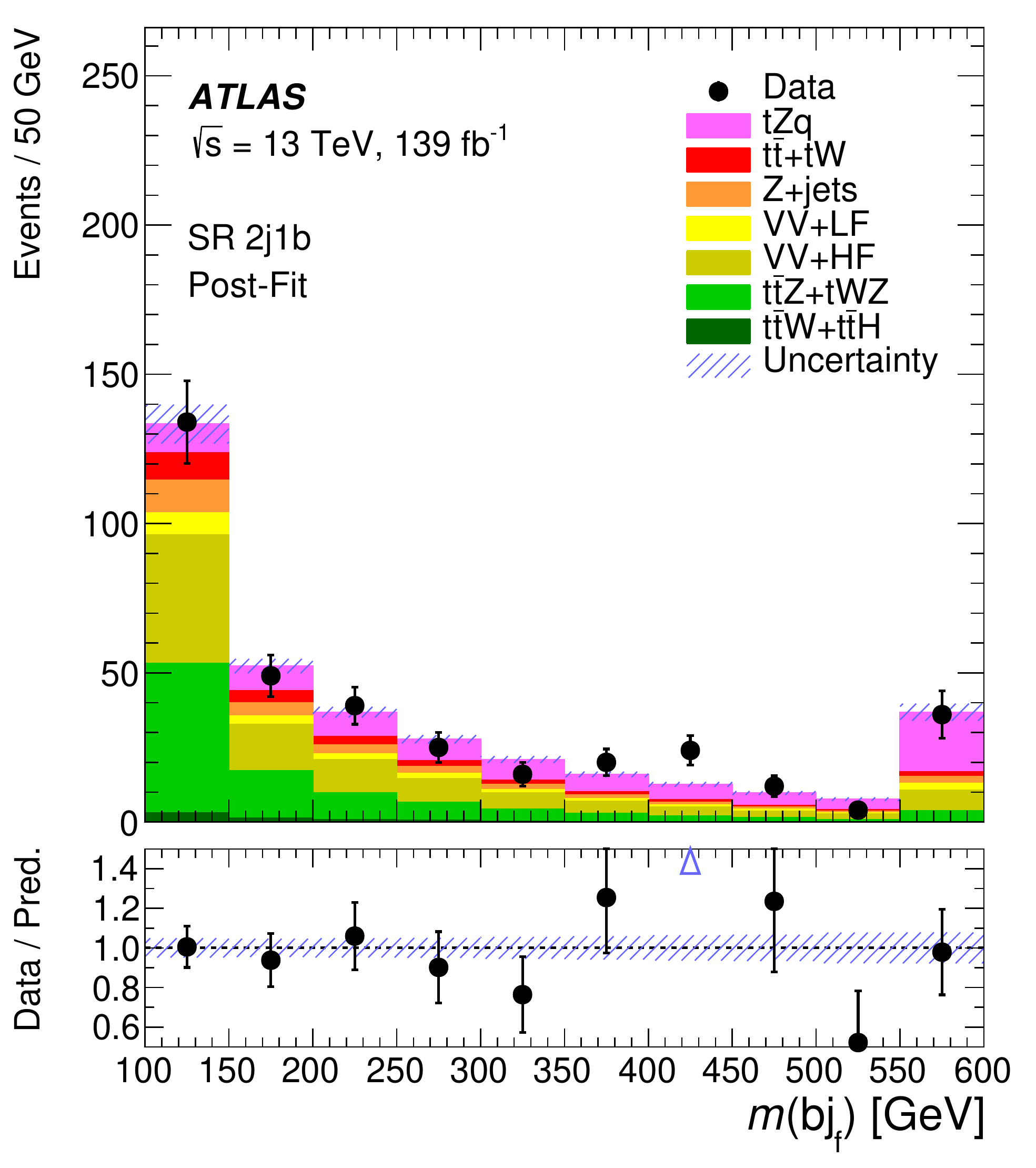}
\caption{}
\end{subfigure}
\begin{subfigure}[b]{.32\textwidth}
\includegraphics[width=\textwidth]{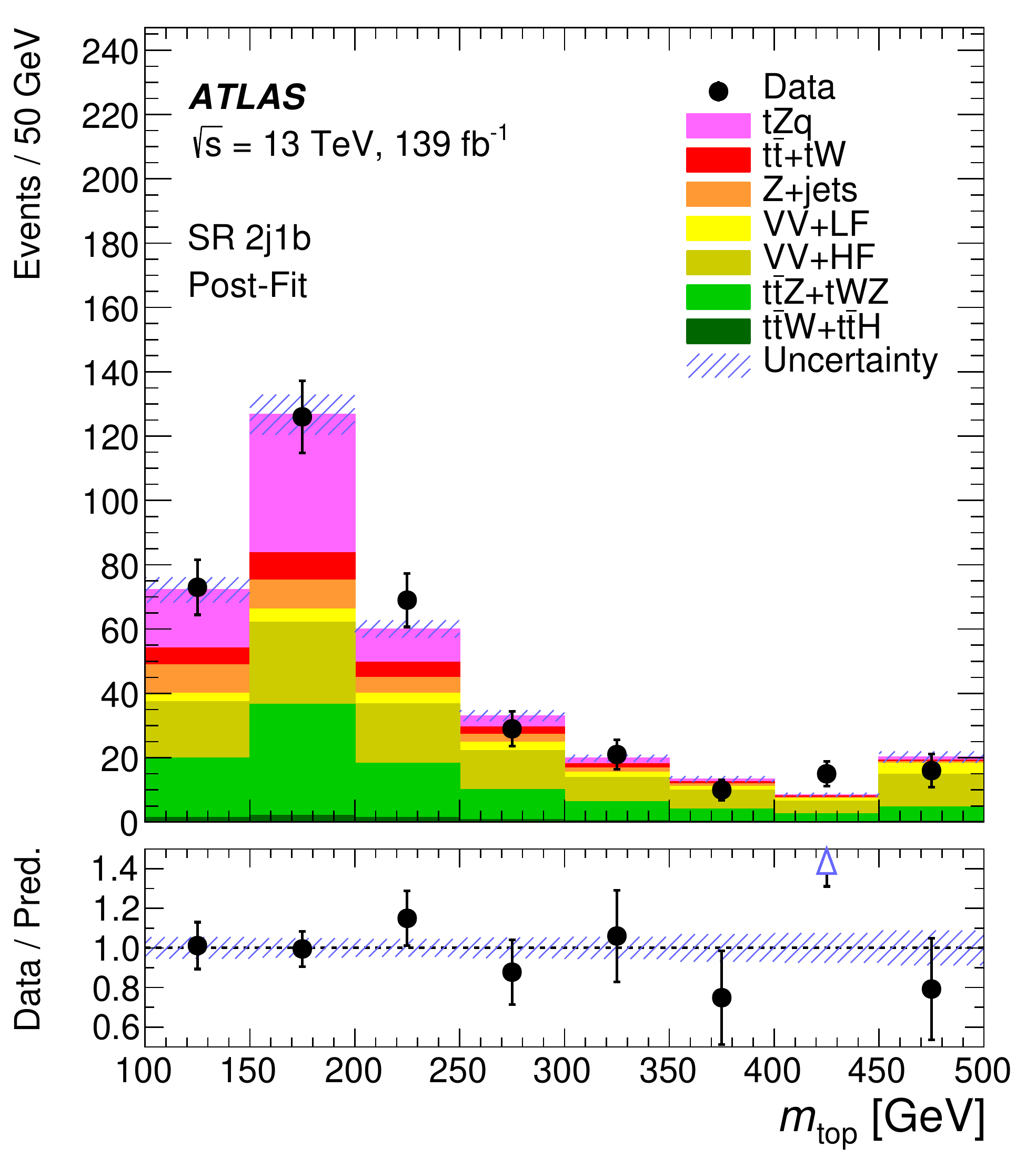}
\caption{}
\end{subfigure}
\begin{subfigure}[b]{.32\textwidth}
\includegraphics[width=\textwidth]{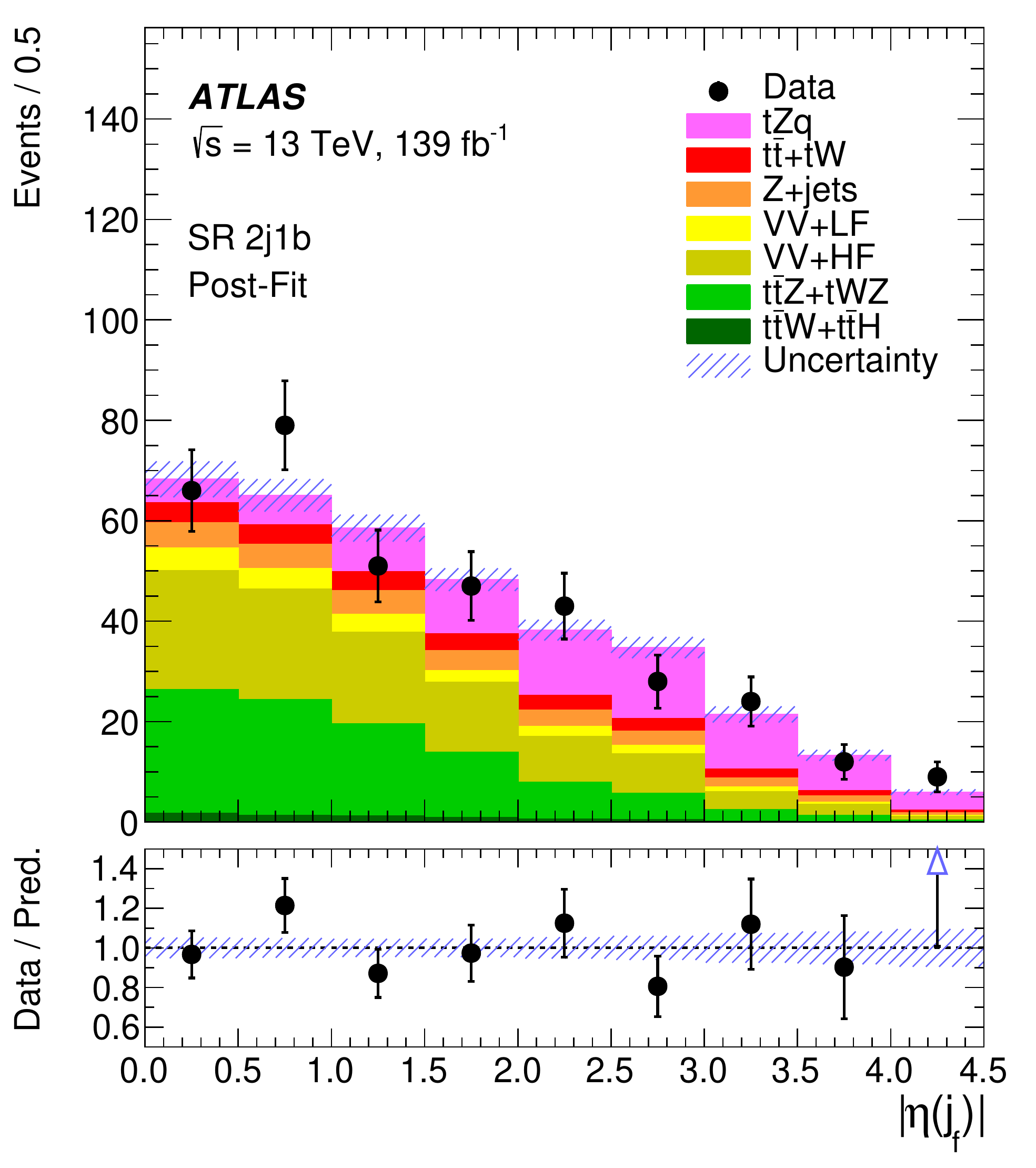}
\caption{}
\end{subfigure} \\
\begin{subfigure}[b]{.32\textwidth}
\includegraphics[width=\textwidth]{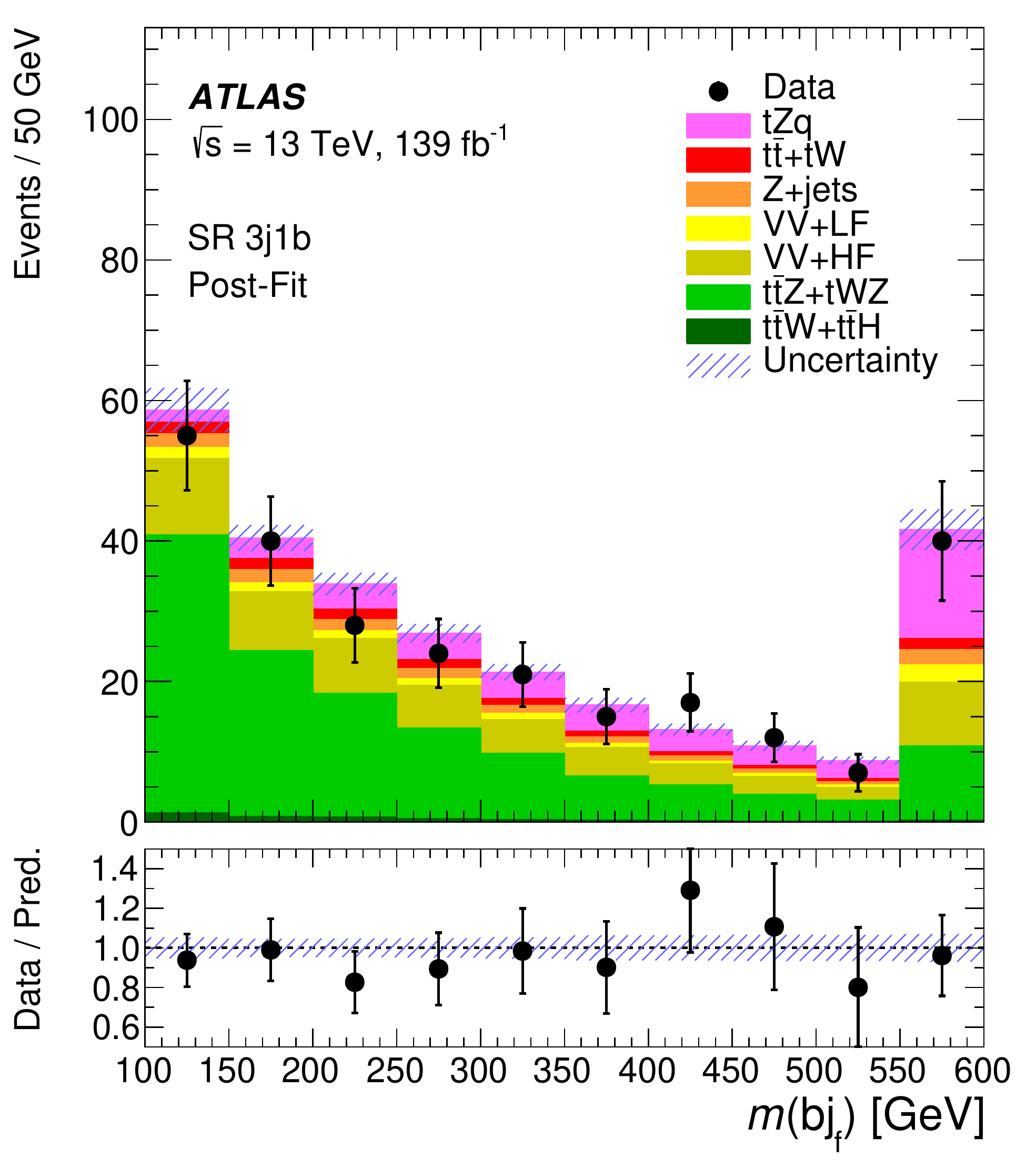}
\caption{}
\end{subfigure}
\begin{subfigure}[b]{.32\textwidth}
\includegraphics[width=\textwidth]{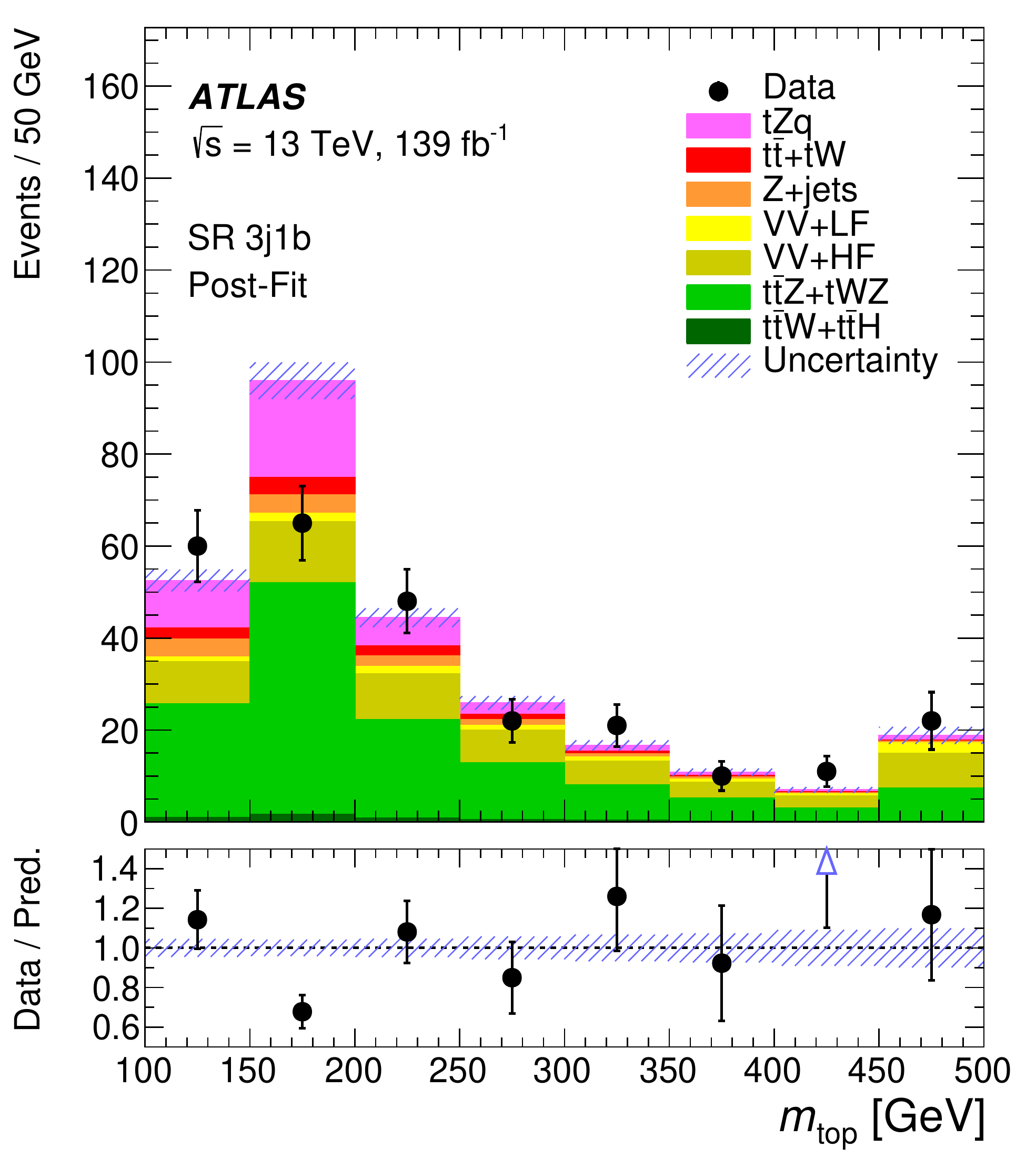}
\caption{}
\end{subfigure}
\begin{subfigure}[b]{.32\textwidth}
\includegraphics[width=\textwidth]{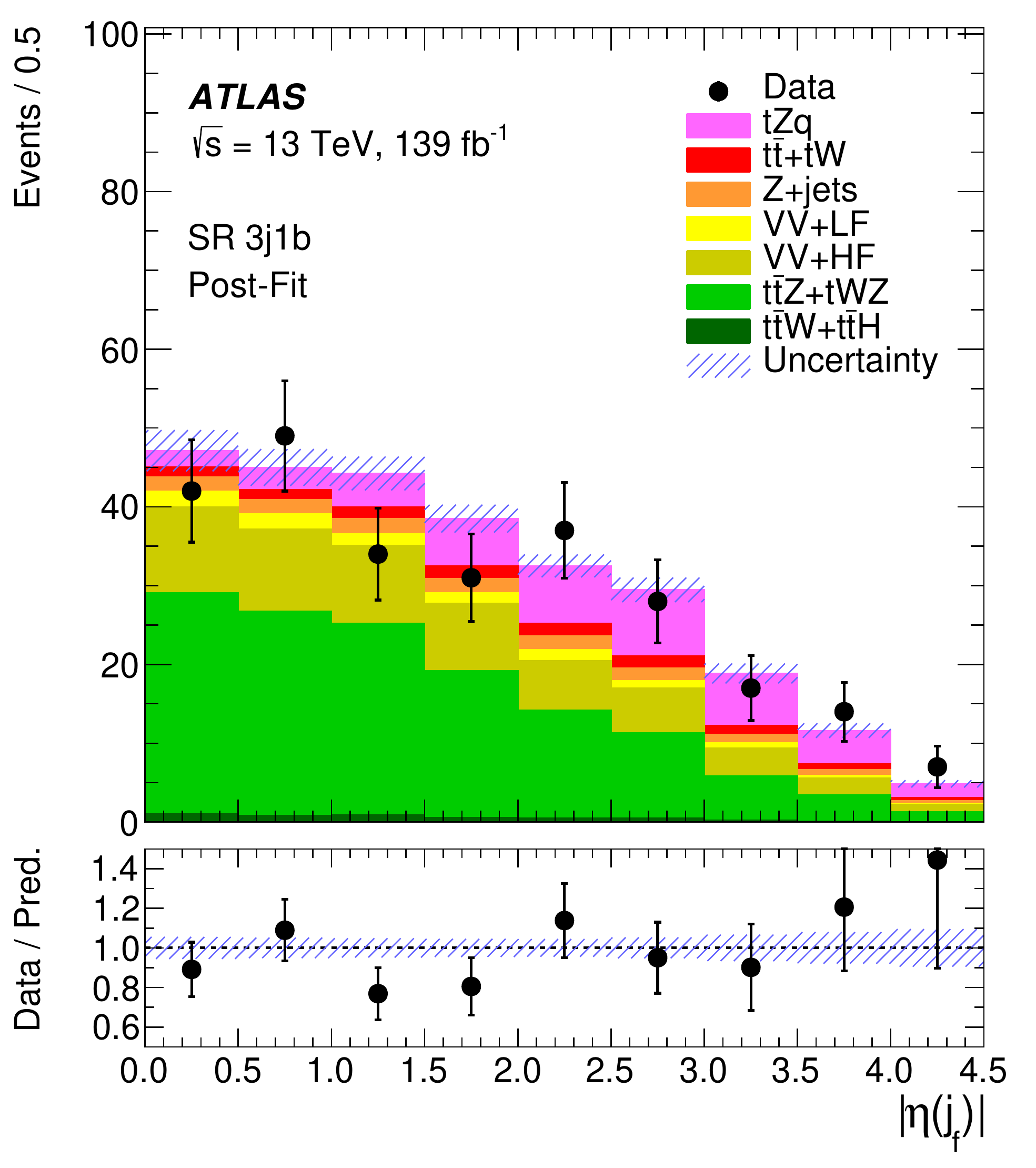}
\caption{}
\end{subfigure}
\caption{Comparison between data and prediction (\enquote{Pred.}) for the three most discriminating NN input variables after the fit to data (\enquote{Post-Fit}) under the signal-plus-background hypothesis. The variables displayed are: (a) and (d) the invariant mass of the \Pqb-jet and the untagged jet, (b) and (e) the mass of the reconstructed top quark, and (c) and (f) the absolute value of the $\eta$ of the \jf jet, shown in the SR 2j1b (top row) and SR 3j1b (bottom row).
The jet denoted \jf is the untagged jet for which the invariant mass of this untagged jet and the \Pqb-tagged jet is the largest.
The uncertainty band includes both the statistical and systematic uncertainties as obtained by the fit.
The rightmost bin includes overflow events. The lower panels show the ratios of the data to the prediction.
The open triangles indicate points that are outside the vertical range of the figure.
}
\label{fig:inputvars}
\end{figure}

The post-fit distributions of the three variables with the highest
discrimination power are shown in \Cref{fig:inputvars} for the two SRs. Good agreement between data and prediction is observed.
 
\begin{table}[t!]
\caption{Variables used as input to the neural network in SR 2j1b and SR 3j1b.
The ranking of the variables in each of the SRs is given in the 2$^\text{nd}$ and 3$^\text{rd}$ columns, respectively. The untagged jet is denoted \jf. When two untagged jets are selected, \jf (\jr) refers to the one for which the invariant mass of this untagged jet and the \Pqb-tagged jet is the largest (smallest).
The \btag score indicates whether the \bjet would also satisfy a tighter \btag requirement corresponding to a working point with an efficiency of 60\% instead of 70\%.}
\label{tab:inputvars}
\begin{center}
\begin{tabular}{l S[table-format=2.0]S[table-format=2.0] p{10.0cm}}
\toprule
Variable & \multicolumn{2}{c}{Rank} & Definition\\
& {SR 2j1b} & {SR 3j1b} & \\
\midrule
\mbj                    &  1 &  1 & (Largest) invariant mass of the \Pqb-jet and the untagged jet(s) \\
\mtrec                   &  2 &  2 & Reconstructed top-quark mass \\
$\abs{\eta(\jf)}$       &  3 &  3 & Absolute value of the $\eta$ of the \jf jet\\
$m_{\text{T}}(\Pl,\MET)$&  4 &  4 & Transverse mass of the \PW boson \\
\btag score             &  5 & 11 & \btag score of the \Pqb-jet \\
$H_{\text{T}}$          &  6 &{--}& Scalar sum of the \pT of the leptons and jets in the event \\
$q(\Pl_{\PW})$          &  7 &  8 & Electric charge of the lepton from the \PW-boson decay\\
$\abs{\eta(\Pl_{\PW})}$ &  8 & 12 & Absolute value of the $\eta$ of the lepton from the \PW-boson decay\\
$\pT(\PW)$              &  9 & 15 & \pT of the reconstructed \PW boson \\
$\pT(\Pl_{\PW})$        & 10 & 14 & \pT of the lepton from the \PW-boson decay\\
$m(\Pl\Pl)$             & 11 &{--}& Mass of the reconstructed \PZ boson\\
$\abs{\eta(\PZ)}$       & 12 & 13 & Absolute value of the $\eta$ of the reconstructed \PZ boson \\
$\Delta R(\jf, \PZ)$    & 13 &  7 &  $\Delta R$ between the \jf jet and the reconstructed \PZ boson \\
$\MET$                  & 14 &{--}& Missing transverse momentum \\
$\pT(\jf)$              & 15 & 10 & \pT of the \jf jet\\
$\abs{\eta(\jr)}$       &{--}&  5 & Absolute value of the $\eta$ of the \jr jet \\
$\pT(\PZ)$              &{--}&  6 & \pT of the reconstructed \PZ boson \\
$\pT(\jr)$              &{--}&  9 & \pT of the \jr jet \\
\bottomrule
\end{tabular}
\end{center}
\end{table}

\section{Systematic uncertainties}
\label{sec:sys}
 
Systematic uncertainties in the signal acceptance
and in the normalisation of the individual backgrounds,
as well as uncertainties in the shape of the fitted distributions,
are taken into account. These are treated as correlated between the different regions, unless stated otherwise.
The uncertainties can be classified into the following categories:
 
\paragraph{Reconstruction efficiency and calibration uncertainties:}
Systematic uncertainties affecting the reconstruction efficiency and energy calibration
of electrons, muons and jets are propagated through the analysis,
each contributing at the percent level to the signal and background uncertainties before the fit.
 
The differences between the electron (muon) trigger, reconstruction and selection efficiencies in data and those in MC simulation are corrected for by scale factors derived from
dedicated \( \PZ \rightarrow \ee \)
(\( \PZ \rightarrow \mumu \)) enriched control samples using a tag-and-probe
method~\cite{EGAM-2018-01-INT,PERF-2015-10}.

The jet energy scale (JES) was derived using information from test-beam data, LHC collision data and simulation, as described in Ref.~\cite{PERF-2016-04}.
The fractional JES uncertainty decreases with the \pT of the reconstructed jet and is rather stable in $\eta$. 
It has various components according to the factors it accounts for and
the different steps used to compute it. 
The impact of the uncertainty in the jet energy resolution is also evaluated.

The \btag efficiencies and mistagging
rate are measured in data using the same methods as
described in Refs.~\cite{FTAG-2018-01-INT,ATLAS-CONF-2018-001,ATLAS-CONF-2018-006}, with the systematic
uncertainties due to \btag efficiency and the mistagging rates
calculated separately.
The impact of the uncertainties on the \btag calibration is evaluated separately
for \Pqb-, \Pqc- and light-jets in the MC samples.
 
The uncertainty in \MET due to a possible miscalibration
of the soft-track component of the \MET is derived from data--MC comparisons
of the \pT balance between the hard and soft \MET components~\cite{PERF-2016-07}. 
The uncertainty associated with the leptons and jets is propagated from the
corresponding uncertainties in the energy/momentum scales and resolutions, and
is classified together with the uncertainty associated with the corresponding objects.
 
\paragraph{Signal and background modelling:}
The systematic uncertainties due to MC modelling of the signal and the \ttbar and \ttZ backgrounds are estimated by comparing different MC generators and by varying the parameters
associated with the renormalisation and factorisation scales, and additional
radiation.
 
For the signal MC simulation, the effects of the systematic uncertainty in the renormalisation and factorisation scales, which are set equal to
each other, and in the amount of additional radiation are calculated simultaneously and are referred to as \enquote{\tZq QCD radiation}.
This is done by increasing the scales by a factor of two and using the \enquote{Var3cDown} parameter of the A14 tune,
and by decreasing the scales by a factor of 0.5 combined with the A14 tune using \enquote{Var3cUp}~\cite{ATL-PHYS-PUB-2014-021}.
The PDF uncertainty in the signal MC prediction is also taken into account by calculating the RMS of the 100 replicas of the \textsc{NNPDF30\_nlo\_as\_0118\_nf\_4} PDF set
following the PDF4LHC prescription~\cite{Butterworth:2015oua}.
The PDF uncertainty shape variations are very small and are therefore neglected.
 
The effect of changing the MC generator for \ttbar events is included as the \ttbar NLO matching systematic uncertainty by taking the difference between the \POWHEGBOX and \MGMCatNLO predictions. The impact of changing the parton shower and hadronisation model is evaluated by comparing samples generated with \POWHEGBOX interfaced with either  \PYTHIAV{8} or \HERWIGV{7}, as described in Ref.~\cite{ATL-PHYS-PUB-2016-004}.
Systematic uncertainties related to the scale and additional radiation are also included using the samples described in~\cref{sec:datamc}. 
 
The effect of changing the MC generator for \ttZ events is included as the \ttZ modelling systematic uncertainty by taking the difference between the \MGMCatNLO and \SHERPA predictions.
Uncertainties related to the choice of renormalisation and factorisation scales, as well as initial-state radiation are also included.

The uncertainty associated with the modelling of diboson events is assessed by comparing the predictions of the \SHERPA and \POWHEGBOX generators. This uncertainty is treated as uncorrelated for the \VVLF and \VVHF components.
 
To account for the possible differences in the distribution shape of \Zjets events originating from sources other than semileptonic \Pqb-decays,
a systematic uncertainty in the shape of the \Zjets distribution is applied.
The systematic uncertainty is constructed by comparing the distribution shapes of MC events from different sources separated according to the origin
of the fake lepton in the event (e.g.\ heavy-flavour decay, photon conversion, etc.).
This uncertainty is not included for \ttbar + \tW since the non-prompt-lepton contributions in CRs and SRs have the same composition and are fully dominated by semileptonic \Pqb-decays.

\paragraph{Background rate uncertainty:}
 
For \ttZ and \tWZ productions, a cross-section uncertainty of 12\% is used~\cite{Florian},
correlated among the two processes.
For \VVLF production, the normalisation uncertainty is taken to be 20\%~\cite{STDM-2018-03}.
The uncertainty for \VVHF production is 30\%~\cite{HIGG-2016-29}. The diboson production uncertainties are taken to be uncorrelated.
In addition, modelling uncertainties are also used in the fit, as mentioned above.
For \ttW and \ttH a 15\% systematic uncertainty in the normalisation is used~\cite{Florian}, correlated among the two processes.
 
For the backgrounds that are unconstrained in the fit (\Zjets and \ttbar + \tW),
an additional normalisation uncertainty of 15\% for \Zjets and 7\% for \ttbar + \tW,
uncorrelated for each region, is included to allow for differences between the regions.
These values correspond to the largest statistical uncertainty in the predicted yields for the \Zjets and \ttbar + \tW MC samples in the SRs and the relevant CRs.

\paragraph{Luminosity:}
The uncertainty in the combined 2015--2018 integrated luminosity is 1.7\%~\cite{ATLAS-CONF-2019-021}, obtained using the LUCID-2 detector~\cite{LUCID2} for the primary luminosity measurements.
 
\paragraph{Uncertainty in \pileup modelling:}
The uncertainty in \pileup modelling is accounted for by varying the reweighting of the MC samples to the data \pileup conditions
using the uncertainty on the average number of interactions per bunch crossing.

 
 
\section{Results}
\label{sec:results}
A simultaneous binned
maximum-likelihood fit
of the SRs and CRs is performed using MC distributions for both the signal
and background predictions. The templates are binned distributions of
the NN output (\ONN) for the SRs and \ttZ CRs;
$m_{\text{T}}(\ell, \MET)$ for the diboson CR,
to provide separation between diboson and \Zjets events in this region; and the total event yield in the \ttbar CRs.
For the \ttZ CR templates, the NN trained in a given SR is evaluated on the events selected in the corresponding CR\@.
In this case, the top-quark reconstruction is performed using the reconstructed \PW boson and \bjet combination that has the invariant mass closest to \mtop.
To measure the \tZq cross-section the normalisation of the
signal template is extracted from the fit.
 
Nuisance parameters are included in the
fit to account for each systematic variation described in \cref{sec:sys}.
When variations of the shape of the distributions are not statistically significant, only the effect on the normalisation is taken into account in the assessment of the systematic uncertainty.
To quantify the statistical significance of the fit and its resulting power
to reject the background-only hypothesis, a test statistic is constructed using a
profile likelihood ratio. 
 
\begin{figure}[t!]
\centering
\begin{subfigure}[b]{.45\textwidth}
\includegraphics[width=\textwidth]{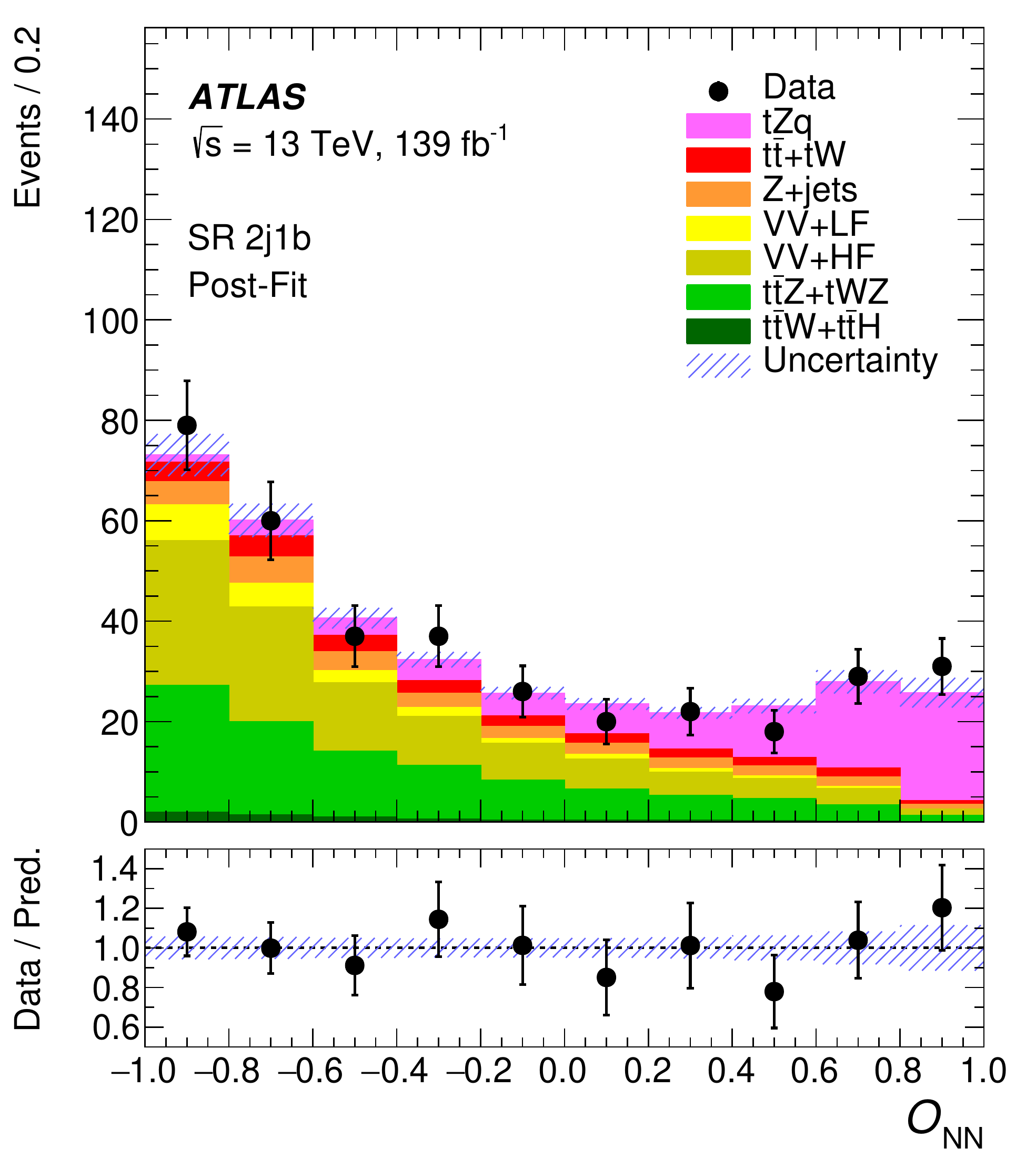}
\caption{}
\end{subfigure}
\begin{subfigure}[b]{.45\textwidth}
\includegraphics[width=\textwidth]{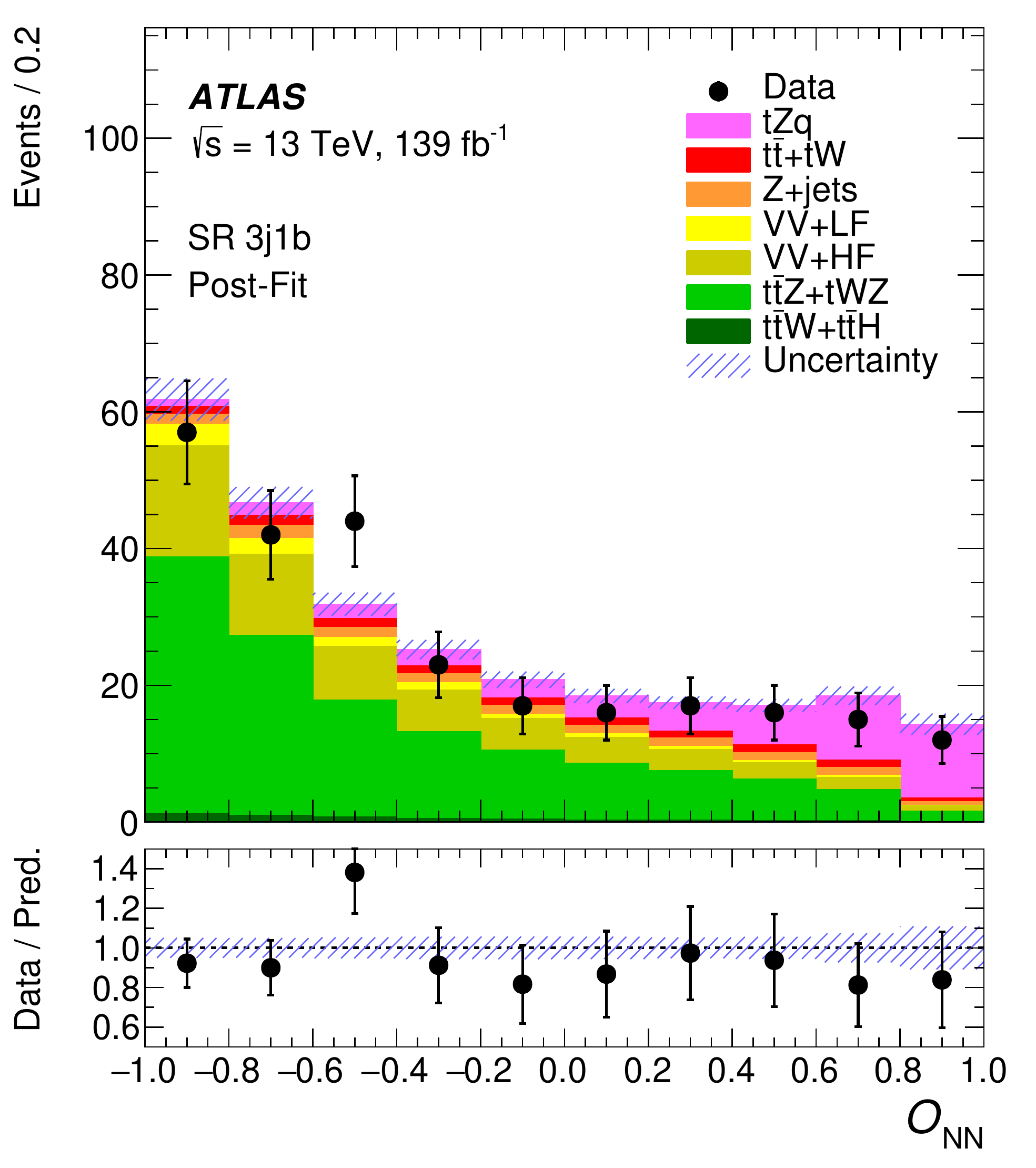}
\caption{}
\end{subfigure}
\caption{Comparison between data and prediction (\enquote{Pred.}) after the fit to data (\enquote{Post-Fit}) under the signal-plus-background hypothesis for the fitted distributions of the neural network output \ONN in the SRs (a) 2j1b and (b) 3j1b.
The uncertainty band includes both the statistical and systematic uncertainties as obtained by the fit. The lower panels show the ratios of the data to the prediction.
}
\label{fig:fitvarsSR}
\end{figure}

\begin{figure}[htbp]
\centering
\begin{subfigure}[b]{.32\textwidth}
\includegraphics[width=\textwidth]{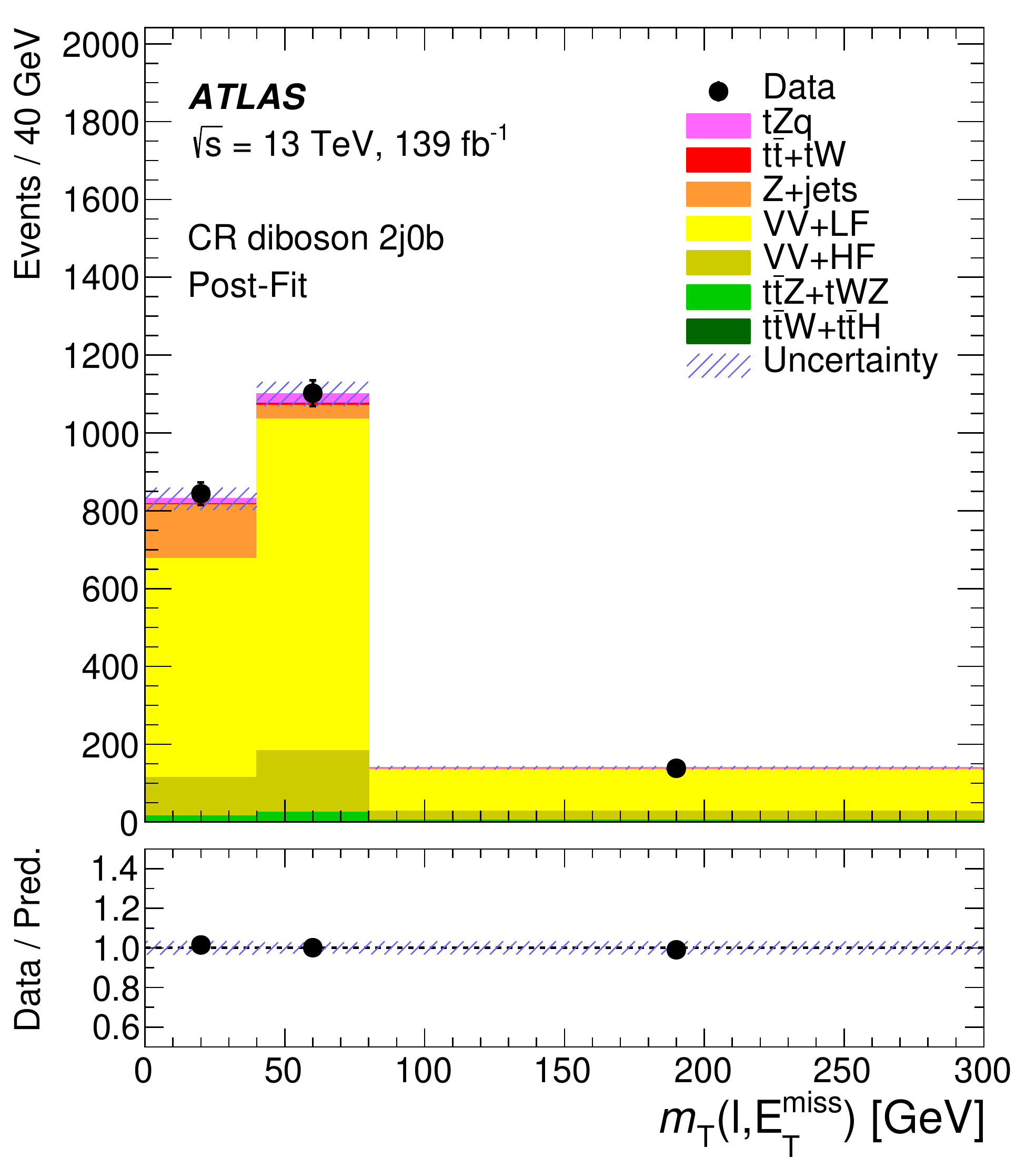}
\caption{}
\end{subfigure}
\begin{subfigure}[b]{.32\textwidth}
\includegraphics[width=\textwidth]{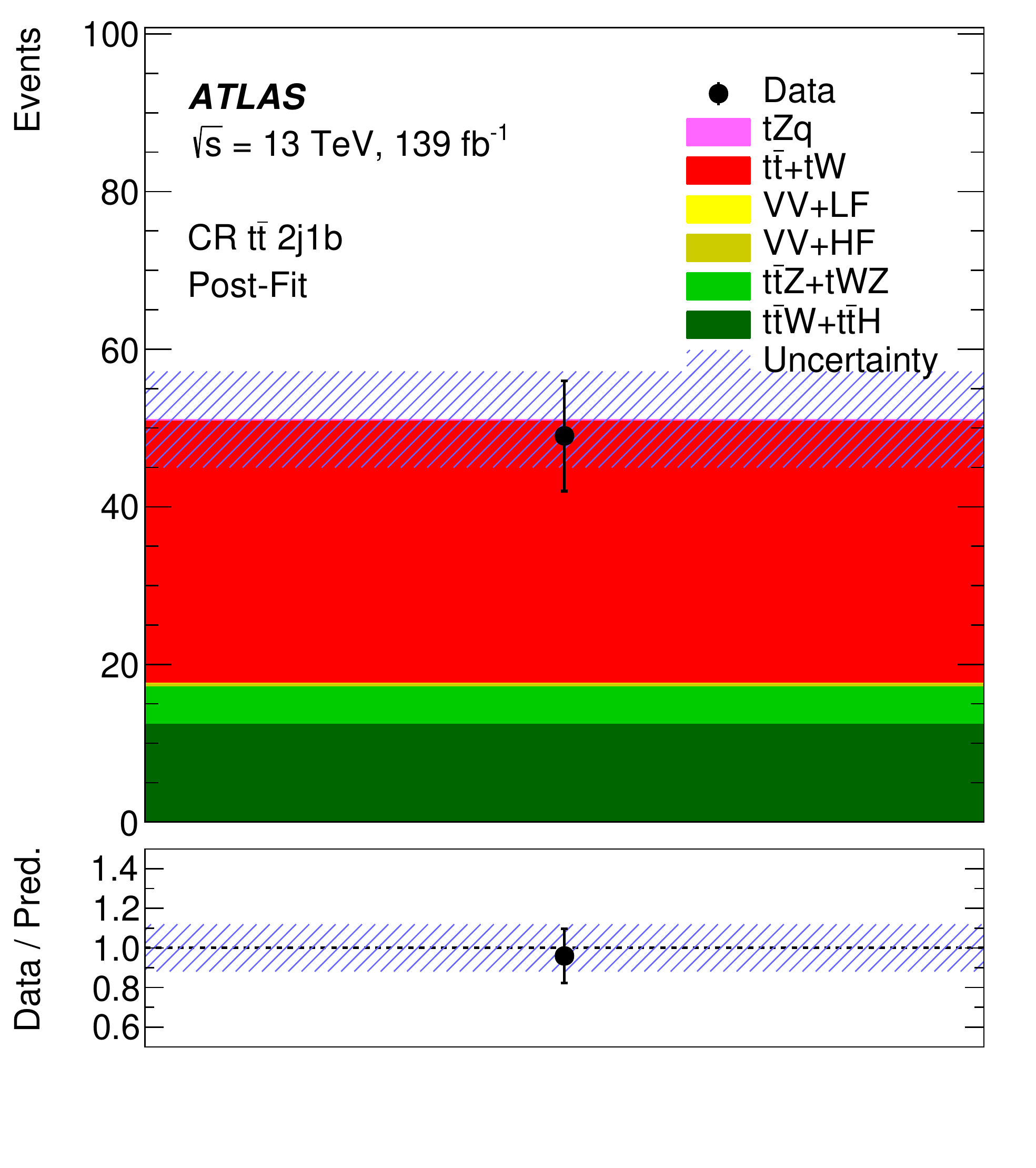}
\caption{}
\end{subfigure}
\begin{subfigure}[b]{.32\textwidth}
\includegraphics[width=\textwidth]{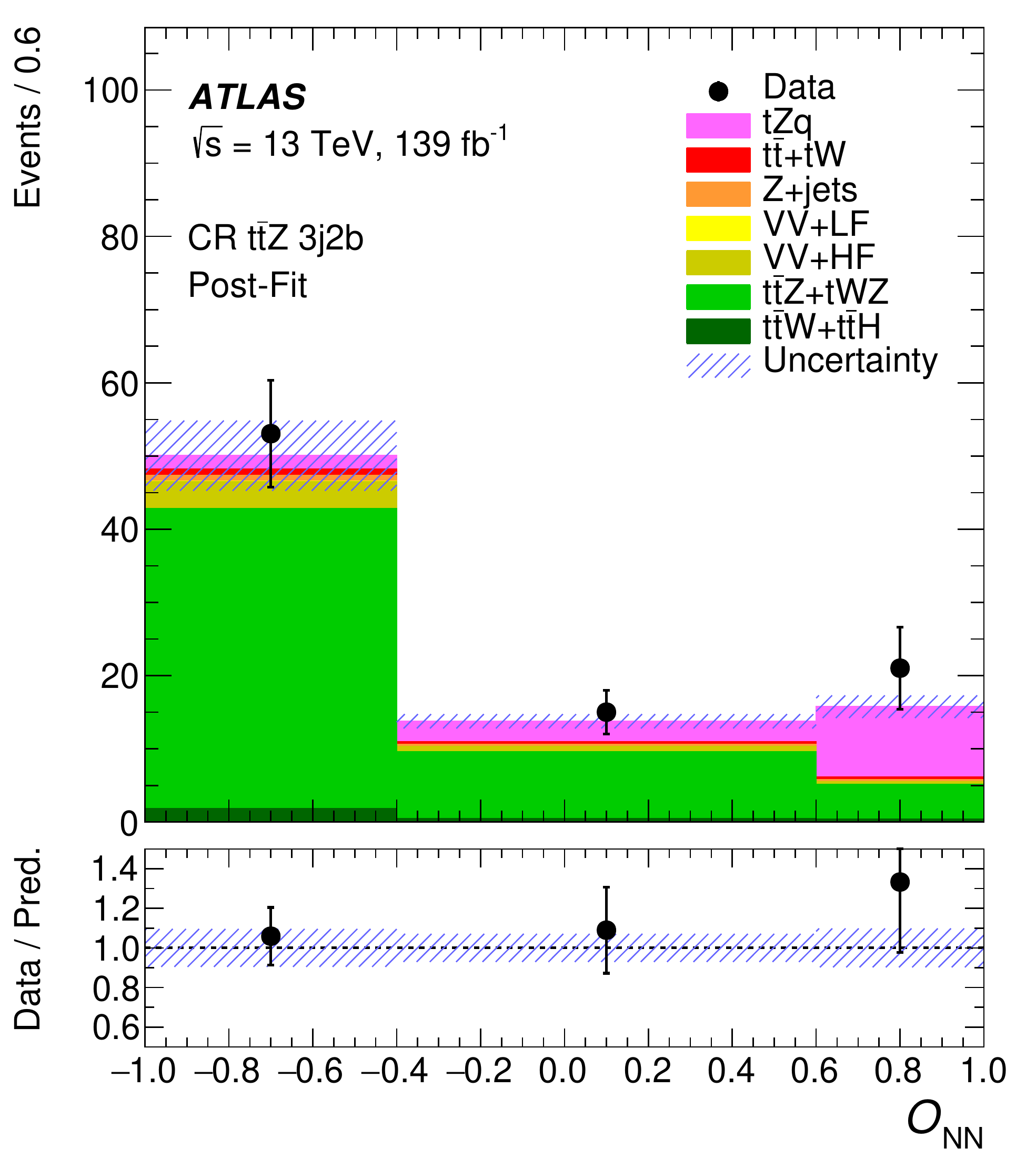}
\caption{}
\end{subfigure} \\
\begin{subfigure}[b]{.32\textwidth}
\includegraphics[width=\textwidth]{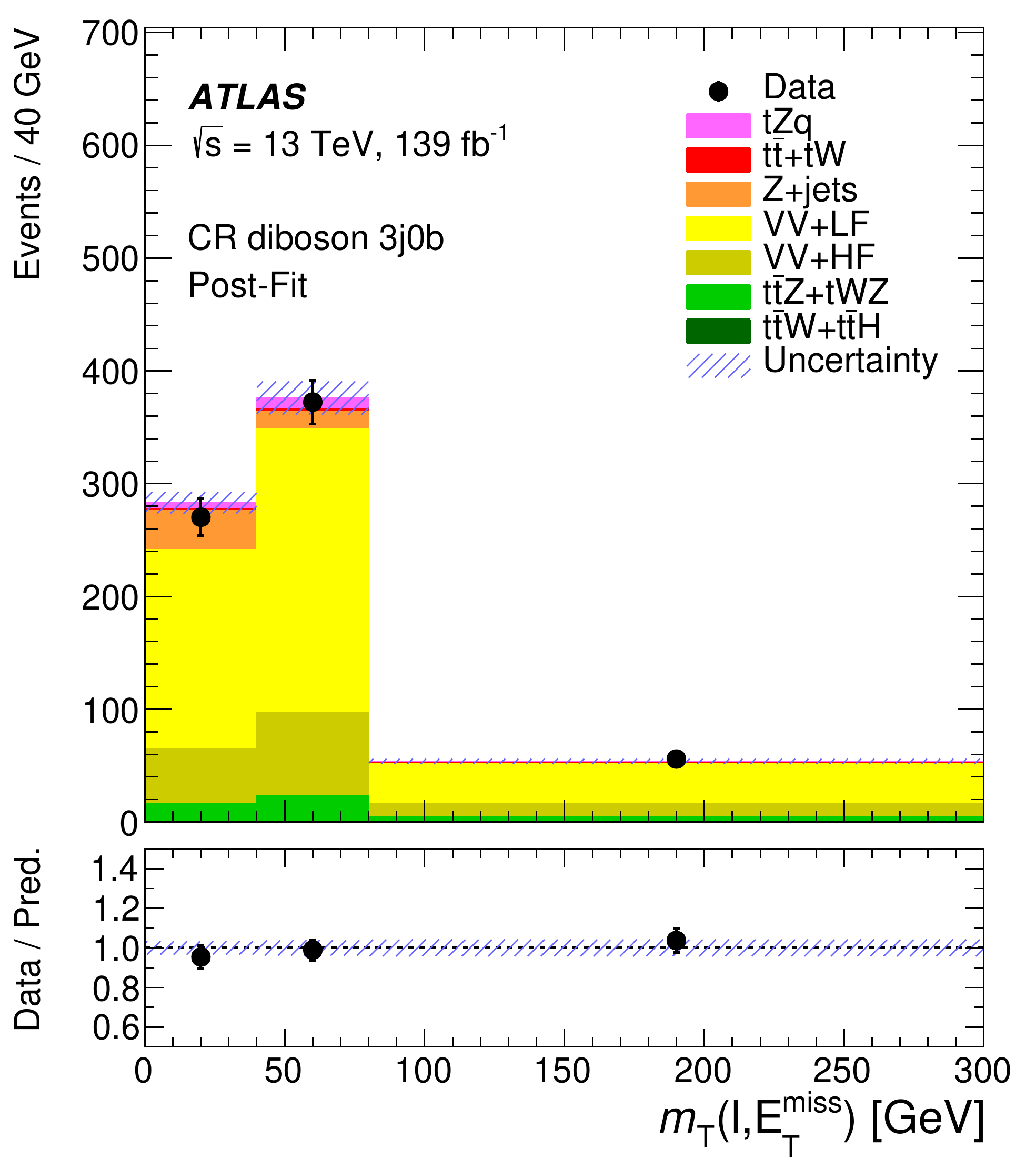}
\caption{}
\end{subfigure}
\begin{subfigure}[b]{.32\textwidth}
\includegraphics[width=\textwidth]{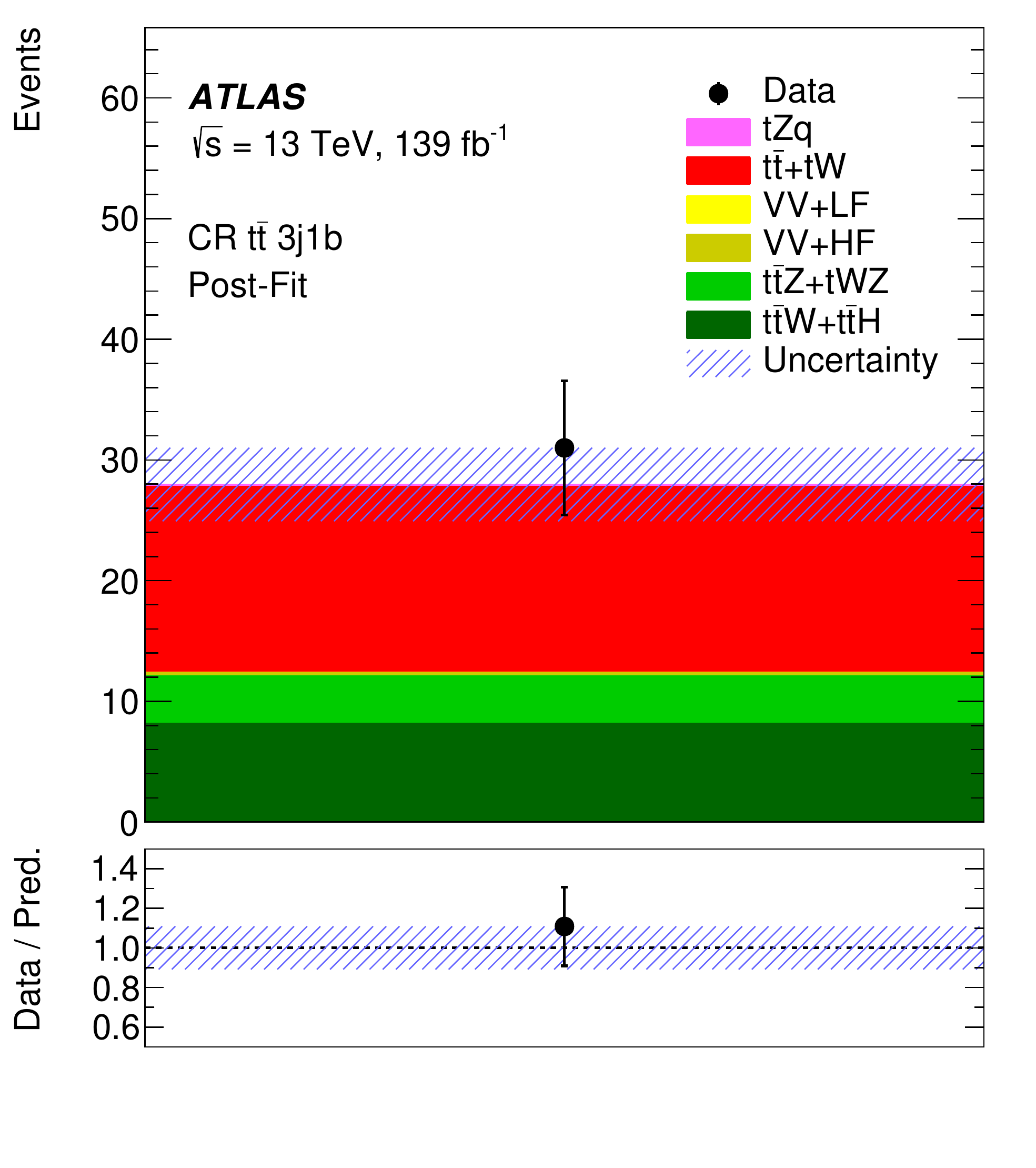}
\caption{}
\end{subfigure}
\begin{subfigure}[b]{.32\textwidth}
\includegraphics[width=\textwidth]{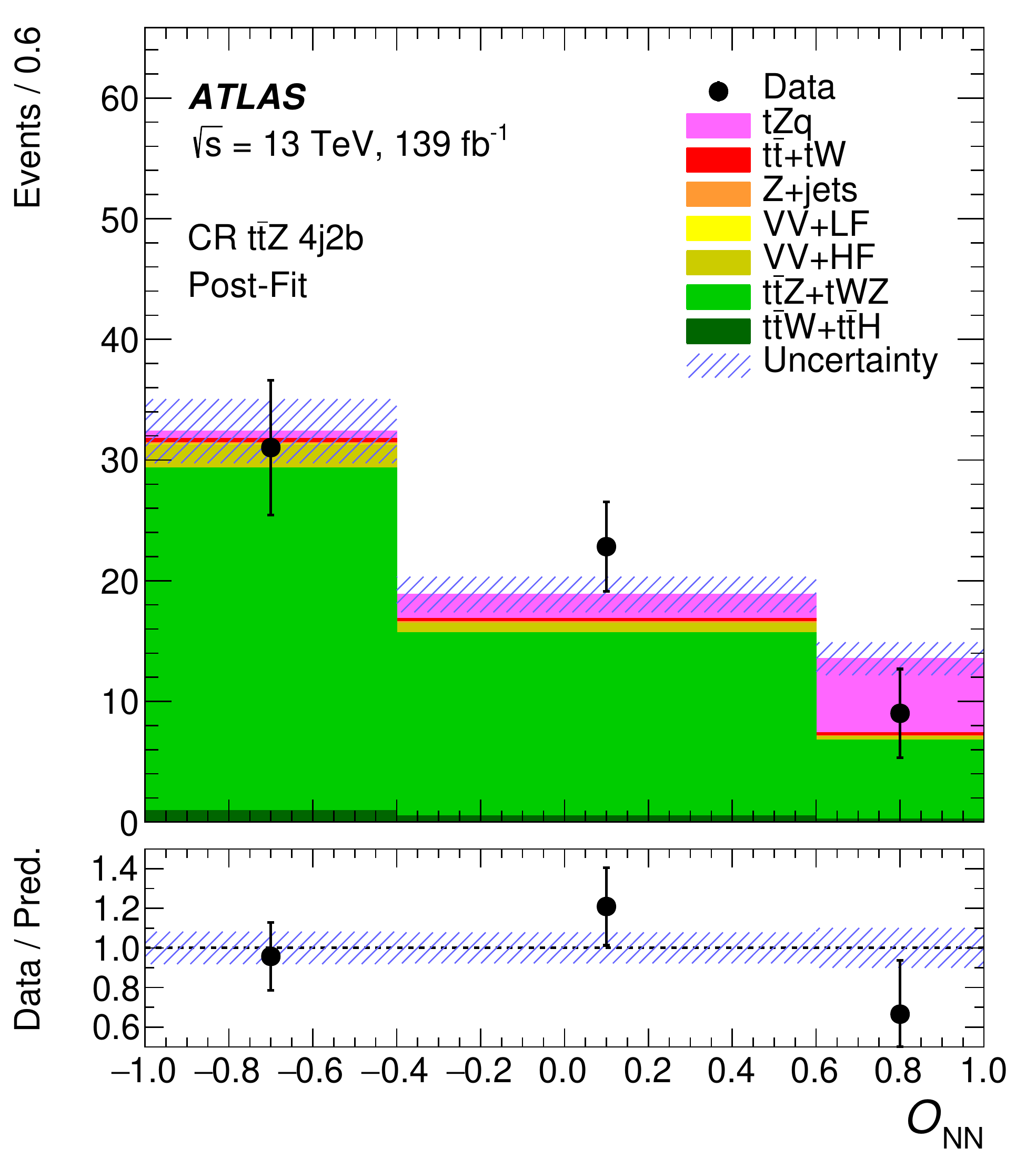}
\caption{}
\end{subfigure}
\caption{Comparison between data and prediction (\enquote{Pred.}) after the fit to data (\enquote{Post-Fit}) under the signal-plus-background hypothesis for the fitted distributions in the CRs.
The fitted distributions are: (a) and (d) the $m_{\text{T}}(\Pl,\MET)$ distribution in the diboson CRs, (b) and (e) the event yields in the \ttbar CRs, and (c) and (f) the \ONN distribution in the \ttZ CRs.
The uncertainty band includes both the statistical and systematic uncertainties as obtained by the fit. The lower panels show the ratios of the data to the prediction.
}
\label{fig:fitvarsCR}
\end{figure}

\Cref{fig:fitvarsSR,fig:fitvarsCR} show the comparison between data and post-fit background and signal distributions, in the SRs and CRs, respectively. 
The numbers of fitted signal and background events compared with the data are shown in \cref{tab:yields}.
The normalisation of the unconstrained fit parameters agrees with the SM predictions.
Validation regions are defined to further check the level of agreement between data and simulation. Details are given in~\cref{app:VR}.
 
The results of the fit yield a \tZq production cross-section, including non-resonant dilepton pairs with $m_{\ell^+\ell^-} > 30\,\text{GeV}$, of \tZxsect,
assuming $\mtop = \SI{172.5}{\GeV}$,
corresponding to a total uncertainty of $\pm$14\%.
The SM cross-section for this process is $\SI[parse-numbers=false]{102^{+5}_{-2}}{\fb}$.
The statistical uncertainty of 13\% in the measurement, which is dominant for this result, is obtained by performing the fit after fixing all nuisance parameters to their post-fit values.
The systematic uncertainty is computed by subtracting in quadrature the statistical
component of the uncertainty from the total.
The impact of the systematic uncertainties on the \tZq cross-section, broken down into major categories, is summarised in \cref{tab:sys}.
 
The statistical significance of obtaining a signal at least as large
as that observed in the data if no signal were present is calculated
using the test statistic in the asymptotic approximation~\cite{Cowan:2010js}.
Both the expected and observed significances are well above five standard deviations.
 
\begin{figure}[t!]
\centering
\begin{subfigure}[b]{.45\textwidth}
\includegraphics[width=\textwidth]{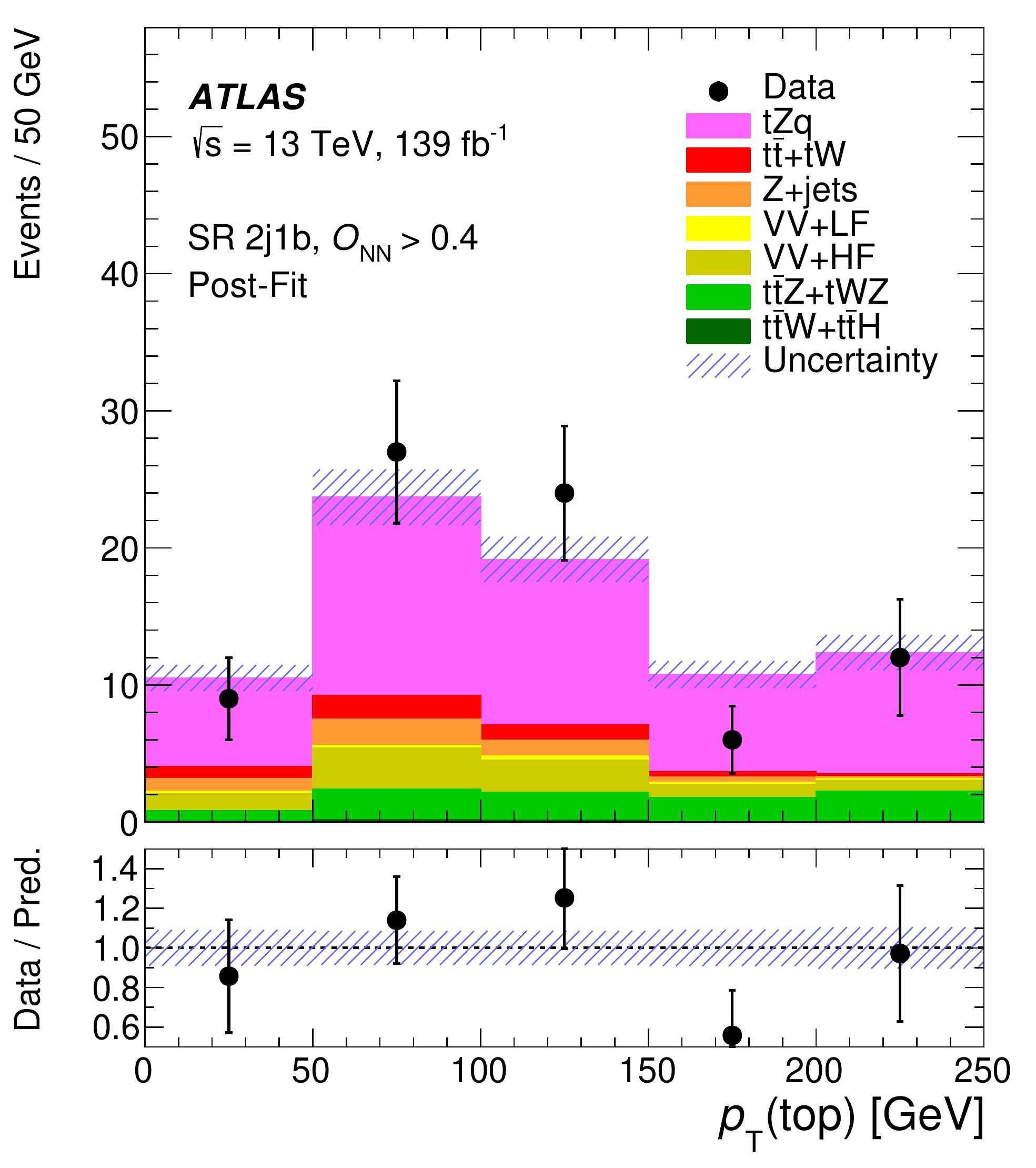}
\caption{}
\end{subfigure}
\begin{subfigure}[b]{.45\textwidth}
\includegraphics[width=\textwidth]{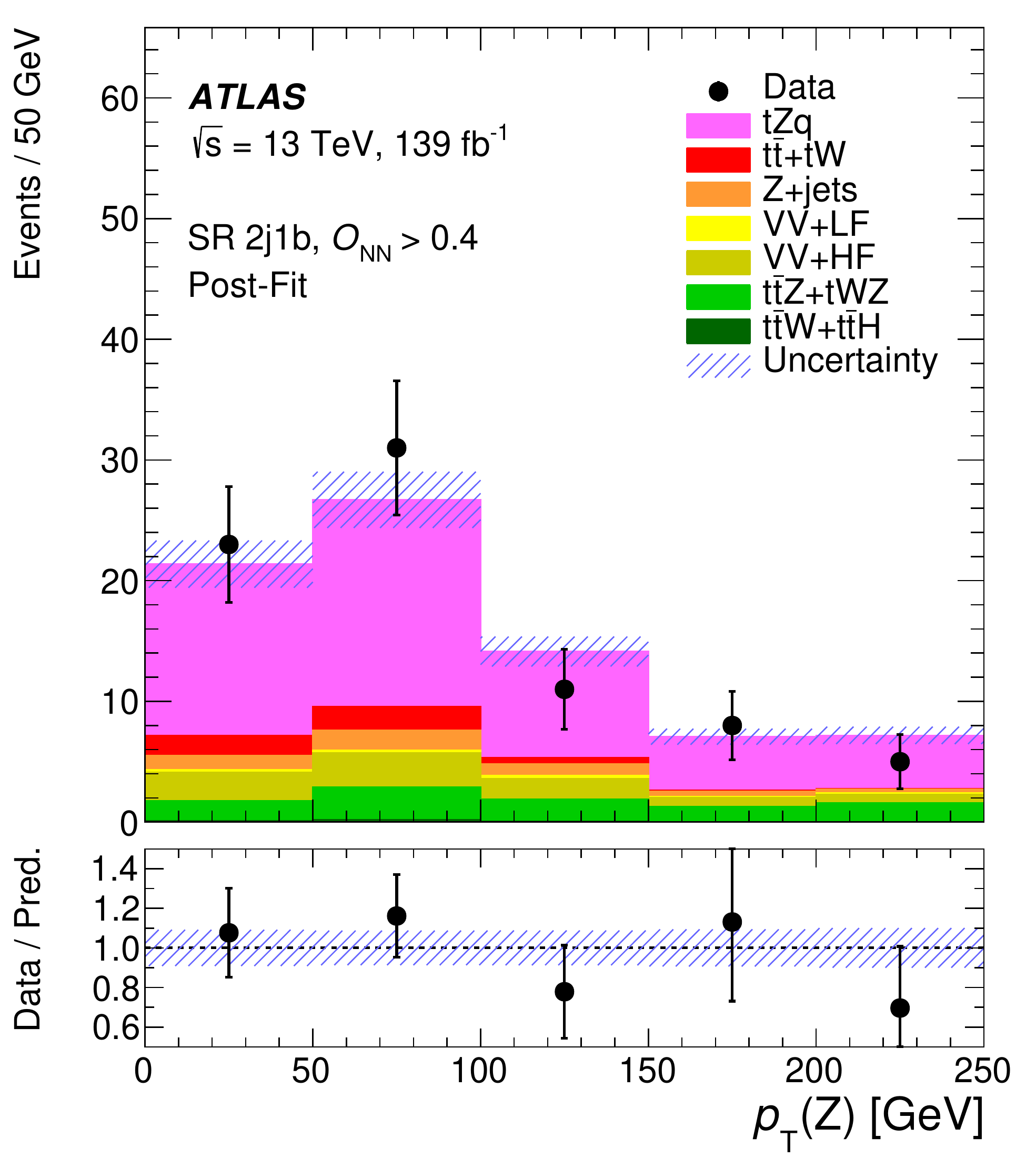}
\caption{}
\end{subfigure}
\caption{Comparison between data and prediction (\enquote{Pred.}) after the fit to data (\enquote{Post-Fit}) under the signal-plus-background hypothesis for the reconstructed \pT of
(a) the top quark and (b) the \PZ boson in the SR 2j1b, for events with $\ONN > 0.4$.
The uncertainty band includes both the statistical and systematic uncertainties as obtained by the fit.
The rightmost bin includes overflow events. The lower panels show the ratios of the data to the prediction.
}
\label{fig:moneyplots}
\end{figure}

The distributions of the reconstructed \pT of the top quark and of the \PZ boson in the SR 2j1b for events with high \ONN score (\ONN $>$ 0.4) are shown in~\Cref{fig:moneyplots}. Good agreement is observed between data and the prediction.
 
\newcommand*{\numRF}[3]{\tablenum[table-format=#1, round-mode=figures, round-precision=#2]{#3}}
\newcommand*{\numRP}[3]{\tablenum[table-format=#1, round-mode=places, round-precision=#2]{#3}}
\begin{table}[htbp]
\caption{
Predicted and observed yields in each of the analysis regions considered.
The signal and background predictions are shown after the fit to data.
The quoted uncertainties include the statistical and systematic uncertainties of the yields,
computed taking into account correlations among nuisance parameters and among processes.
}
\label{tab:yields}
\begin{center}
\begin{tabular}{l
*4{S[table-format=4.1]@{$\,\pm\,$}S[table-format=2.1]}
}
\toprule
& \multicolumn{2}{c}{SR 2j1b}                          & \multicolumn{2}{c}{CR diboson 2j0b}                 & \multicolumn{2}{c}{CR \ttbar 2j1b}                     & \multicolumn{2}{c}{CR \ttZ 3j2b} \\
\midrule
\tZq         & 79 & 11    & 53.1 & 7.5 & 0.2 & 0.1                   & 12.9 & 2.0 \\
\ttbar + \tW & 23.8 & 4.8 & 13.7 & 2.7 & 33.3 & 6.3                  & 1.7 & 0.3 \\
\Zjets       & 28 & 13    & 181 & 82   & \multicolumn{2}{c}{$< 0.1$}      & 1.4 & 0.6 \\
\VVLF        & 19.7 & 7.9 & 2000 & 100 & \multicolumn{2}{c}{$< 0.1$} & 0.1 & 0.1 \\
\VVHF        & 101 & 22   & 383 & 78   & 0.4 & 0.1                   & 5.2 & 1.7 \\
\ttZ + \tWZ  & 96 & 11    & 63.2 & 7.0 & 4.8 & 0.5                   & 59.3 & 7.1 \\
\ttH + \ttW  & 6.5 & 1.0  & 3.0 & 0.5  & 12.4 & 1.9                  & 2.8 & 0.5 \\
\midrule
Total &
354  & 16  &
2697 & 56  &
51.1 & 6.1 &
83.5 & 6.4 \\
\midrule
Data         & \multicolumn{2}{c}{359} & \multicolumn{2}{c}{2703} & \multicolumn{2}{c}{49} & \multicolumn{2}{c}{92} \\
\bottomrule
\end{tabular}\\
\vspace*{1em}
\begin{tabular}{l
*4{S[table-format=4.1]@{$\,\pm\,$}S[table-format=2.1]}
}
\toprule
& \multicolumn{2}{c}{SR 3j1b}                          & \multicolumn{2}{c}{CR diboson 3j0b}                 & \multicolumn{2}{c}{CR \ttbar 3j1b}                     & \multicolumn{2}{c}{CR \ttZ 4j2b} \\
\midrule
\tZq         & 43.4 & 6.2 & 21.2 & 3.3 & 0.2  & 0.1                  & 8.0  & 1.3 \\
\ttbar + \tW & 11.0 & 2.2 & 6.9  & 1.3 & 15.4 & 3.1                  & 1.0  & 0.2 \\
\Zjets       & 12.8 & 6.0 & 53   & 23  & \multicolumn{2}{c}{$< 0.1$}      & 0.4  & 0.2 \\
\VVLF        & 10.1 & 4.2 & 624  & 53  & \multicolumn{2}{c}{$< 0.1$} & 0.1  & 0.1 \\
\VVHF        & 58   & 17  & 186  & 51  & 0.3  & 0.1                  & 3.4  & 1.0 \\
\ttZ + \tWZ  & 132  & 12  & 61.9 & 6.2 & 3.9  & 0.5                  & 58.1 & 5.3 \\
\ttH + \ttW  & 4.7  & 0.7 & 1.7 & 0.3  & 8.2  & 1.3                  & 2.0  & 0.3 \\
\midrule
Total &
272  & 12  &
955  & 29  &
28.0 & 3.0 &
72.8 & 5.0 \\
\midrule
Data         & \multicolumn{2}{c}{259}  & \multicolumn{2}{c}{949} & \multicolumn{2}{c}{31} & \multicolumn{2}{c}{75} \\
\bottomrule
\end{tabular}
\end{center}
\end{table}
 
\begin{table}[htbp]
\caption{Impact of systematic uncertainties on the \tZq cross-section, broken down into major categories.
For each category the impact is calculated by performing a fit where the nuisance parameters in the group are fixed to their best-fit values,
and then subtracting the resulting uncertainty in the parameter of interest in quadrature from the uncertainty from the nominal fit.
For simplicity, the impact is given as the average of the up and down variations.
Details of the systematic uncertainties are provided in the text.
MC statistics refers to the effect of the limited size of the MC samples.
The total systematic uncertainty is a bit larger than the quadratic sum of the individual contributions due to correlations.
}
\label{tab:sys}
\begin{center}
\sisetup{round-mode=places, round-precision=1, scientific-notation=fixed, fixed-exponent=0}
\begin{tabular}{l
*1{S[table-format=4.1]@{$\,$}S[table-format=2.1]}
}
\toprule
{Uncertainty source}      & {$\Delta\sigma/\sigma$ [\%]} \\
\midrule
 
Prompt-lepton background modelling and normalisation          & 3.26 \\
Jets and \MET reconstruction and calibration                             & 2.01 \\
Lepton reconstruction and calibration                                        & 2.0 \\
Luminosity                & 1.66 \\
Non-prompt-lepton background modelling                              & 1.58 \\
\Pileup modelling                     & 1.20 \\
MC statistics             & 1.04 \\
\tZq modelling (QCD radiation)    & 0.76 \\
\tZq modelling (PDF)                    & 0.65 \\
Jet flavour tagging                        & 0.44 \\
\midrule
Total systematic uncertainty         & 6.96 \\
\midrule
Data statistics &		12.62		\\
\ttbar + \tW and \Zjets normalisation &		2.13		\\
\midrule
Total statistical uncertainty                    & 12.88 \\
 
\bottomrule
\end{tabular}
\end{center}
\end{table}

\FloatBarrier
 
\section{Conclusion}
\label{sec:conclusion}
 
The cross-section for \tZq production,
including non-resonant dilepton pairs with $m_{\ell^+\ell^-} > 30\,\text{GeV}$,
is measured. The analysis uses
\lumi of proton--proton collision data collected
by the ATLAS experiment at the LHC between 2015 and 2018
at a centre-of-mass energy of \SI{13}{\TeV}.
The result of this measurement is \tZxsect, assuming a top-quark mass of \SI{172.5}{\GeV},
corresponding to a total uncertainty of $\pm$14\%.
This result is in good agreement with the SM prediction of $\SI[parse-numbers=false]{102^{+5}_{-2}}{\fb}$.

\section*{Acknowledgements}
 

We thank CERN for the very successful operation of the LHC, as well as the
support staff from our institutions without whom ATLAS could not be
operated efficiently.
 
We acknowledge the support of ANPCyT, Argentina; YerPhI, Armenia; ARC, Australia; BMWFW and FWF, Austria; ANAS, Azerbaijan; SSTC, Belarus; CNPq and FAPESP, Brazil; NSERC, NRC and CFI, Canada; CERN; CONICYT, Chile; CAS, MOST and NSFC, China; COLCIENCIAS, Colombia; MSMT CR, MPO CR and VSC CR, Czech Republic; DNRF and DNSRC, Denmark; IN2P3-CNRS and CEA-DRF/IRFU, France; SRNSFG, Georgia; BMBF, HGF and MPG, Germany; GSRT, Greece; RGC and Hong Kong SAR, China; ISF and Benoziyo Center, Israel; INFN, Italy; MEXT and JSPS, Japan; CNRST, Morocco; NWO, Netherlands; RCN, Norway; MNiSW and NCN, Poland; FCT, Portugal; MNE/IFA, Romania; MES of Russia and NRC KI, Russia Federation; JINR; MESTD, Serbia; MSSR, Slovakia; ARRS and MIZ\v{S}, Slovenia; DST/NRF, South Africa; MINECO, Spain; SRC and Wallenberg Foundation, Sweden; SERI, SNSF and Cantons of Bern and Geneva, Switzerland; MOST, Taiwan; TAEK, Turkey; STFC, United Kingdom; DOE and NSF, United States of America. In addition, individual groups and members have received support from BCKDF, CANARIE, Compute Canada and CRC, Canada; ERC, ERDF, Horizon 2020, Marie Sk{\l}odowska-Curie Actions and COST, European Union; Investissements d'Avenir Labex, Investissements d'Avenir Idex and ANR, France; DFG and AvH Foundation, Germany; Herakleitos, Thales and Aristeia programmes co-financed by EU-ESF and the Greek NSRF, Greece; BSF-NSF and GIF, Israel; CERCA Programme Generalitat de Catalunya and PROMETEO Programme Generalitat Valenciana, Spain; G\"{o}ran Gustafssons Stiftelse, Sweden; The Royal Society and Leverhulme Trust, United Kingdom.
 
The crucial computing support from all WLCG partners is acknowledged gratefully, in particular from CERN, the ATLAS Tier-1 facilities at TRIUMF (Canada), NDGF (Denmark, Norway, Sweden), CC-IN2P3 (France), KIT/GridKA (Germany), INFN-CNAF (Italy), NL-T1 (Netherlands), PIC (Spain), ASGC (Taiwan), RAL (UK) and BNL (USA), the Tier-2 facilities worldwide and large non-WLCG resource providers. Major contributors of computing resources are listed in Ref.~\cite{ATL-GEN-PUB-2016-002}.
 
 
\clearpage
\appendix
\clearpage
 
\part*{Appendix}
\addcontentsline{toc}{part}{Appendix}
 
\section{Validation regions}
\label{app:VR}
 
The normalisation and modelling of the four main backgrounds are
constrained using CRs.
Additional diboson and \ttbar+\ttV validation regions (VRs) are used to confirm the goodness of the background modelling.
These VRs are defined so as to be as close as possible to both the SRs and CRs.
 
The diboson VRs are defined using the same selection as the SRs except that
a \enquote{loose} \Pqb-tag requirement (indicated as \textit{1Lb} in the region name) is imposed that has a \bjet selection
efficiency in the range 85\%--70\%.
The \ttbar +~\ttV VRs are defined using the same selection as the SRs except that
the \OSSF invariant mass requirement is inverted.
 
\begin{figure}[htbp]
\centering
\begin{subfigure}[b]{.45\textwidth}
\includegraphics[width=\textwidth]{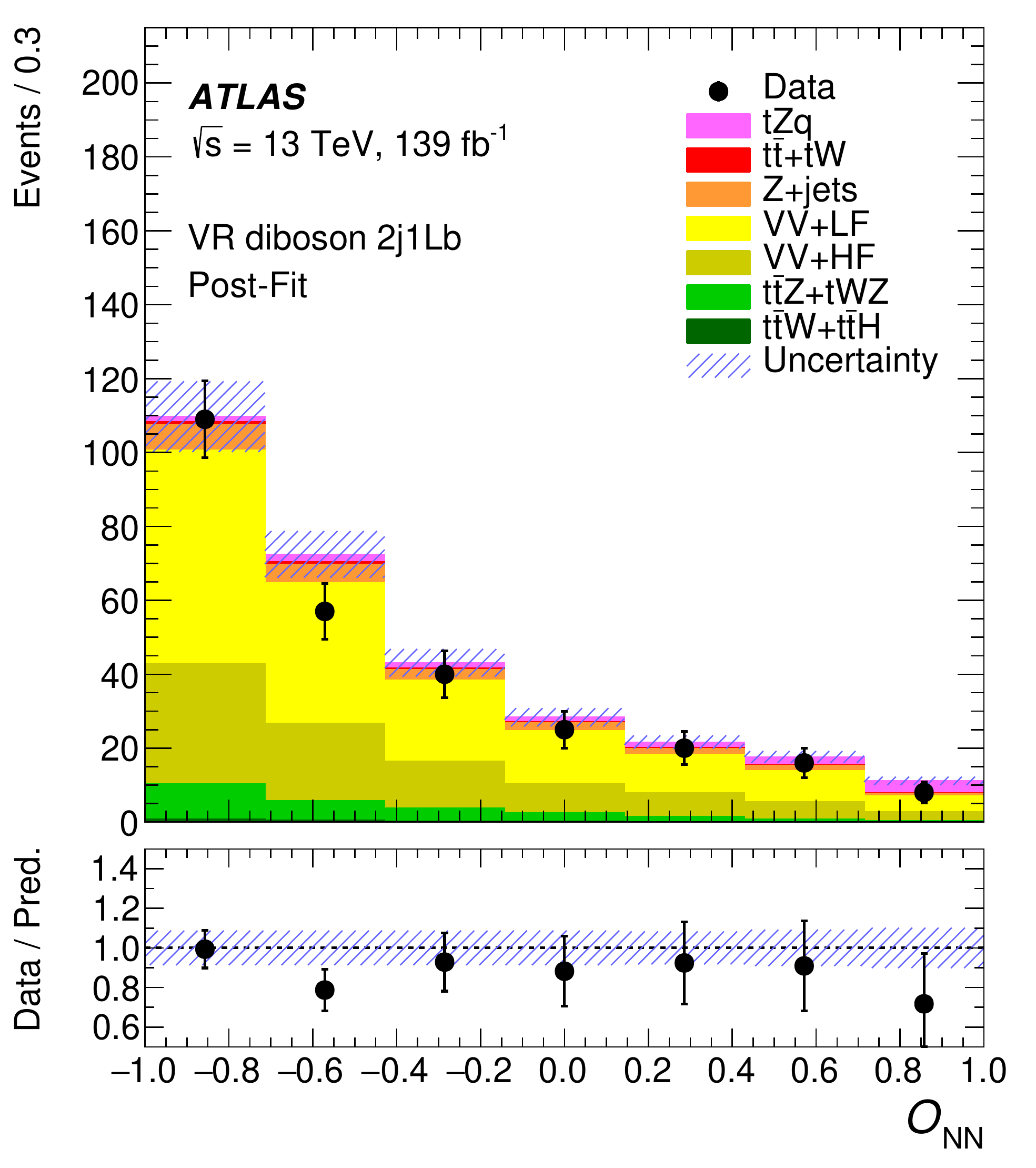}
\caption{}
\end{subfigure}
\begin{subfigure}[b]{.45\textwidth}
\includegraphics[width=\textwidth]{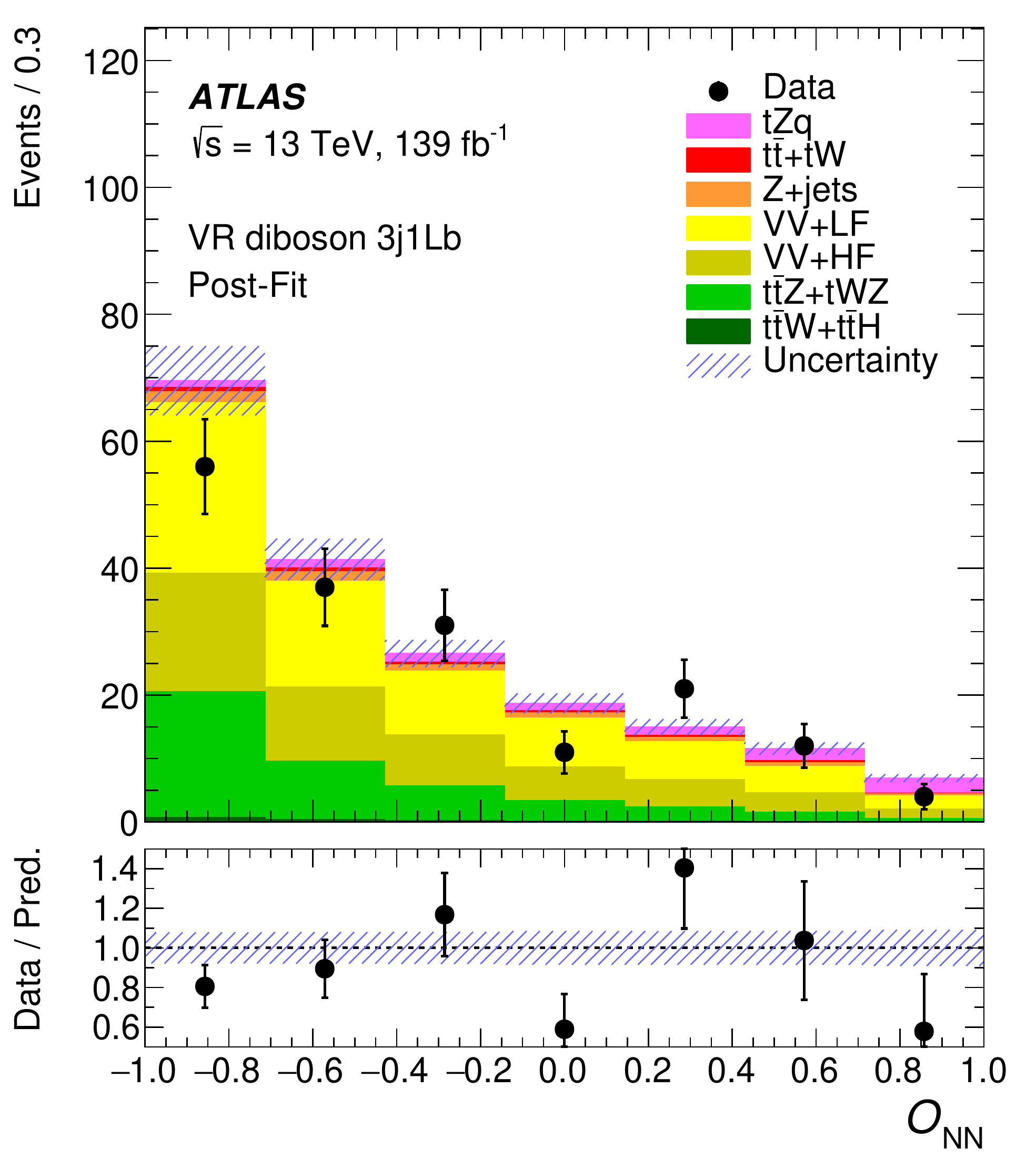}
\caption{}
\end{subfigure} \\
\begin{subfigure}[b]{.45\textwidth}
\includegraphics[width=\textwidth]{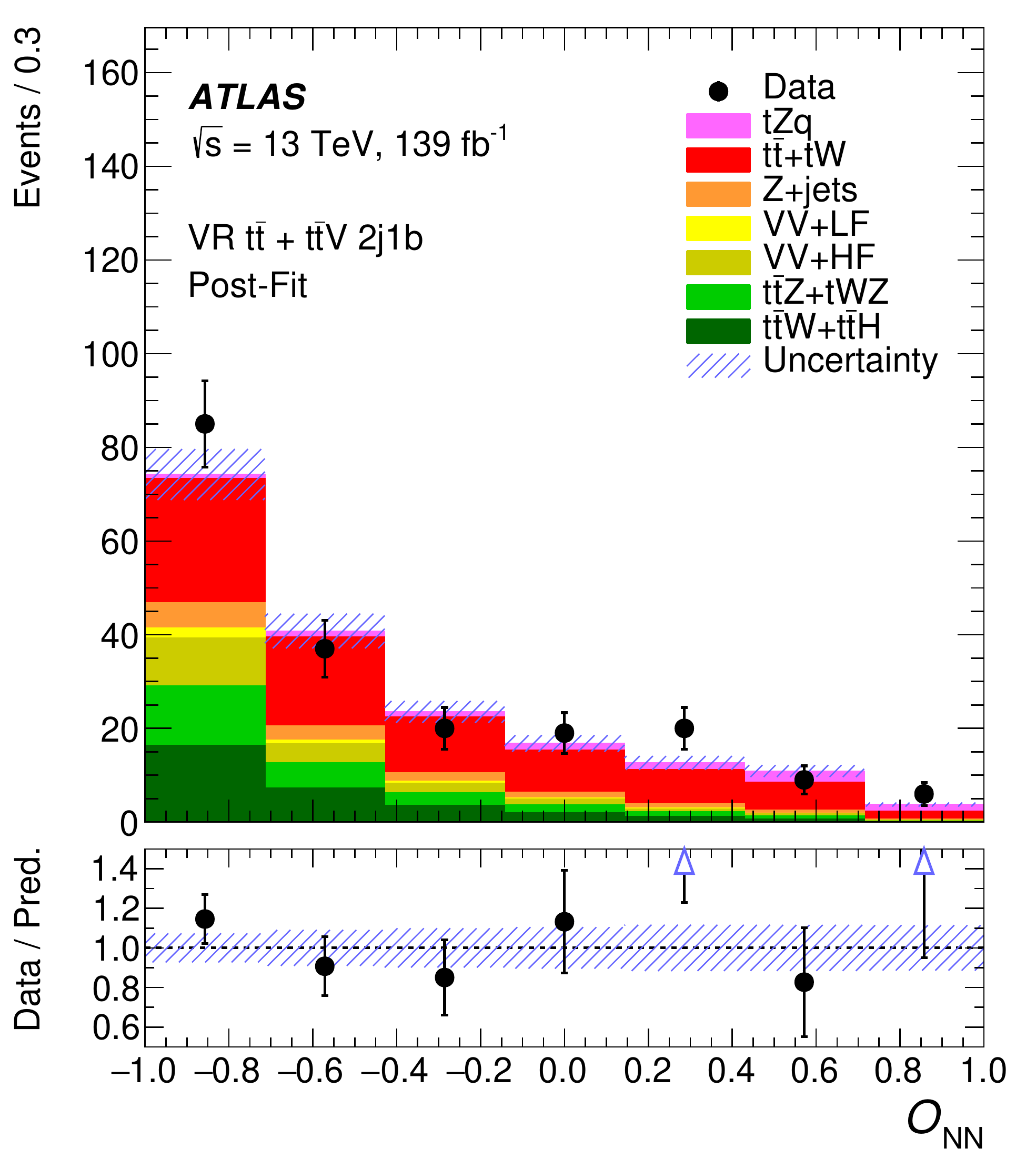}
\caption{}
\end{subfigure}
\begin{subfigure}[b]{.45\textwidth}
\includegraphics[width=\textwidth]{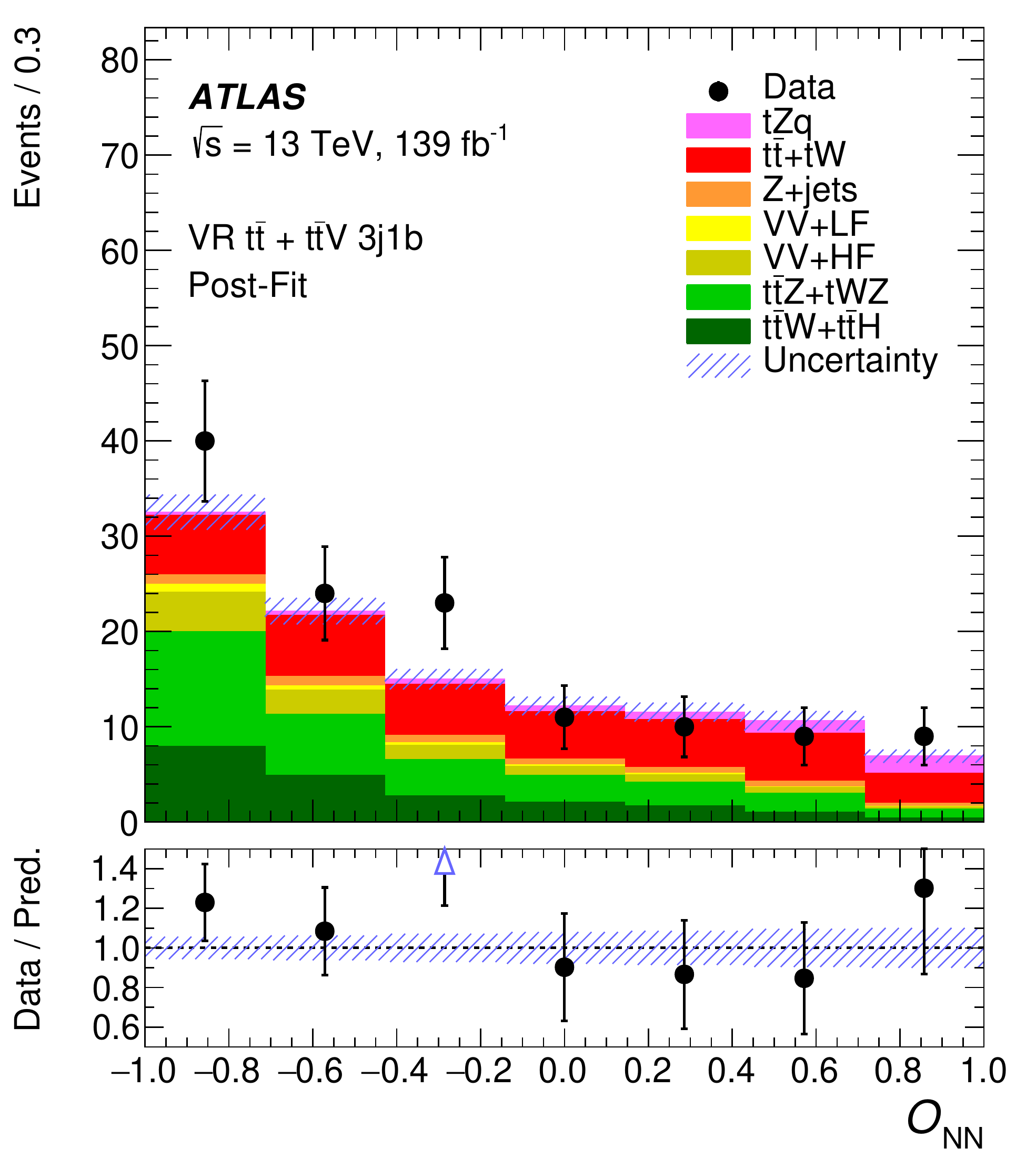}
\caption{}
\end{subfigure}
\caption{Comparison between data and prediction (\enquote{Pred.}) after the fit to data (\enquote{Post-Fit}) under the signal-plus-background hypothesis for the \ONN distributions in the VRs (a) diboson 2j1Lb, (b) diboson 3j1Lb, (c) \ttbar +~\ttV 2j1b and (d) \ttbar +~\ttV 3j1b.
The uncertainty band includes both the statistical and systematic uncertainties as obtained by the fit. The lower panels show the ratios of the data to the prediction.
The open triangles indicate points that are outside the vertical range of the figure.
}
\label{app:fig:vrplots}
\end{figure}

The NN training from each SR is applied to the corresponding VRs. Distributions of the \ONN in the various VRs are shown
in~\Cref{app:fig:vrplots}. Any variables used in the NN training that are undefined in the VRs (e.g.\ due to missing jets) are replaced by a dummy value such that the variable is not used by NeuroBayes during the evaluation.
This is also valid for the variables that have a different range compared to the one that was used in the training, such as the \btag score.

\clearpage
 
\printbibliography
 
\clearpage \input{atlas_authlist}

\end{document}

%% file: atlas_authlist.tex
 
\begin{flushleft}
\hypersetup{urlcolor=black}
{\Large The ATLAS Collaboration}

\bigskip

\AtlasOrcid[0000-0002-6665-4934]{G.~Aad}$^\textrm{\scriptsize 102}$,    
\AtlasOrcid[0000-0002-5888-2734]{B.~Abbott}$^\textrm{\scriptsize 129}$,    
\AtlasOrcid{D.C.~Abbott}$^\textrm{\scriptsize 103}$,    
\AtlasOrcid[0000-0002-2788-3822]{A.~Abed~Abud}$^\textrm{\scriptsize 36}$,    
\AtlasOrcid[0000-0002-1002-1652]{K.~Abeling}$^\textrm{\scriptsize 53}$,    
\AtlasOrcid[0000-0002-2987-4006]{D.K.~Abhayasinghe}$^\textrm{\scriptsize 94}$,    
\AtlasOrcid[0000-0002-8496-9294]{S.H.~Abidi}$^\textrm{\scriptsize 167}$,    
\AtlasOrcid[0000-0002-8279-9324]{O.S.~AbouZeid}$^\textrm{\scriptsize 40}$,    
\AtlasOrcid{N.L.~Abraham}$^\textrm{\scriptsize 156}$,    
\AtlasOrcid[0000-0001-5329-6640]{H.~Abramowicz}$^\textrm{\scriptsize 161}$
% ...
\bigskip
\\

% ...
$^{36}$CERN, Geneva; Switzerland.\\
% ...
$^{40}$Niels Bohr Institute, University of Copenhagen, Copenhagen; Denmark.\\
% ...
$^{53}$II. Physikalisches Institut, Georg-August-Universit\"{a}t G\"ottingen, G\"ottingen; Germany.\\
% ...
$^{94}$Department of Physics, Royal Holloway University of London, Egham; United Kingdom.\\
% ...
$^{102}$CPPM, Aix-Marseille Universit\'e, CNRS/IN2P3, Marseille; France.\\
$^{103}$Department of Physics, University of Massachusetts, Amherst MA; United States of America.\\
% ...
$^{129}$Homer L. Dodge Department of Physics and Astronomy, University of Oklahoma, Norman OK; United States of America.\\
% ...
$^{156}$Department of Physics and Astronomy, University of Sussex, Brighton; United Kingdom.\\
% ...
$^{161}$Raymond and Beverly Sackler School of Physics and Astronomy, Tel Aviv University, Tel Aviv; Israel.\\
% ...
$^{167}$Department of Physics, University of Toronto, Toronto ON; Canada.\\
% ...

\end{flushleft}
